\documentclass[runningheads]{llncs}
\usepackage[T1]{fontenc}
\usepackage[utf8]{inputenc}
\usepackage{microtype}
\usepackage{listings}
\usepackage{mathpartir}
\usepackage{enumitem}
\usepackage{amsmath}
\usepackage{amssymb}
\usepackage{mathtools}
\usepackage{stmaryrd}
\usepackage{array}
\usepackage{thm-restate}
\usepackage[bookmarksdepth=2]{hyperref}
\usepackage{xcolor}

\usepackage{macros}

\AddToHook{cmd/appendix/before}{}

\lstdefinelanguage{source}{
  morekeywords=[1]{let, in, handle, if, then, else},
  morecomment=[l]{\#}
}

\definecolor{target}{RGB}{0, 96, 192}
\definecolor{meta}{RGB}{192, 96, 0}
\lstdefinelanguage{meta}{
  basicstyle=\codestyle,
  keywordstyle=\metakwstyle,
  morekeywords=[1]{let, in, match, with, fresh, fail, if, then, else},
  moredelim={[is][\targetkwstyle]{\{}{\}}},
  morecomment=[s]{(*}{*)},
  commentstyle=\commentstyle
}

\begin{document}
\title{Deciding not to Decide}
\subtitle{Sound and Complete Effect Inference\\
  in the Presence of Higher-Rank Polymorphism}
%
%
\author{Patrycja Balik \and Szymon Jędras \and Piotr Polesiuk}
%
\authorrunning{Balik, Jędras, and Polesiuk}
%
\institute{University of Wrocław \\
  \email{\{pbalik,sjedras,ppolesiuk\}@cs.uni.wroc.pl}}
%
\maketitle              
\begin{abstract}
  Type-and-effect systems help the programmer to organize data and
  computational effects in a program.
  While for traditional type systems expressive variants with sophisticated
  inference algorithms have been developed
  and widely used in programming languages,
  type-and-effect systems did not yet gain widespread adoption.
  One reason for this is that type-and-effect systems are more complex
  and the existing inference algorithms make compromises between
  expressiveness, intuitiveness, and decidability.
  In this work, we present an effect inference algorithm
  for a type-and-effect system with subtyping,
  expressive higher-rank polymorphism,
  and intuitive set-like semantics of effects.
  In order to deal with scoping issues of higher-rank polymorphism,
  we delay solving of effect constraints by transforming them into
  formulae of propositional logic.
  We prove soundness and completeness of our algorithm
  with respect to a declarative type-and-effect system.
  All the presented results have been formalized in the Rocq proof assistant,
  and the algorithm has been successfully implemented
  in a realistic programming language.
\end{abstract}


\sbox{\thmcodeTrType}{\lstinline[language=meta]{tr_type(}}
\sbox{\thmcodesubtype}{\lstinline[language=meta]{subtype(}}
\sbox{\thmcodeinfer}{\lstinline[language=meta]{infer(}}
\sbox{\thmcodesep}{\lstinline[language=meta]{;}}
\sbox{\thmcodebrop}{\lstinline[language=meta]{(}}
\sbox{\thmcodebrcl}{\lstinline[language=meta]{)}}

\section{Introduction}

Type-and-effect systems~\cite{%
  DBLP:conf/popl/JouvelotG91,%
  DBLP:journals/jfp/TalpinJ92}
permit tracking information not just about the data in a program, but also the
allowed behavior of computations. This information is especially important in
languages with advanced control mechanisms, such as those featuring algebraic
effect handlers~\cite{DBLP:journals/corr/PlotkinP13}. Ideally,
a programmer-facing type-and-effect system should be ergonomic, expressive, and
intuitive.

In terms of ergonomics, one obstacle is that effect information tends to be
quite large. Fortunately, this can be remedied with effect reconstruction,
where the effect information omitted by the programmer can be automatically
inferred.
As for expressiveness, a usable effect system almost certainly needs a form
of effect polymorphism.
Finally, intuitiveness requires that the representation of effects matches
the user's understanding of effect information.
Many effect systems described in the literature~\cite{%
  DBLP:conf/popl/JouvelotG91,%
  DBLP:journals/jfp/TalpinJ92,%
  nielson2000type,%
  DBLP:journals/pacmpl/ZhangM19,%
  DBLP:journals/pacmpl/BrachthauserSO20,%
  DBLP:journals/pacmpl/MadsenP20,%
  DBLP:journals/pacmpl/BiernackiPPS20%
} represent this information as a set of possible behaviors.
This design choice leads to both an
elegant theory and intuitive semantics of types with effects for the
programmer, but is not always easy to implement in practice.

A common approach is to use effect rows~\cite{%
  DBLP:conf/tldi/LindleyC12,%
  DBLP:conf/icfp/HillerstromL16,%
  DBLP:journals/corr/Leijen14,%
  DBLP:conf/sfp/IkemoriCML22,%
  DBLP:conf/esop/VilhenaP23}
to approximate sets of effects. Effect rows can be thought of as a list of
behaviors, potentially terminated by a polymorphic row variable.
This technique can be easily combined with the Hindley-Milner algorithm, which
is the golden standard for type reconstruction, and grants some additional
expressiveness in the form of ML-polymorphism.
The key advantage of using rows is that unification is decidable and
unitary~\cite{DBLP:conf/lics/Wand87,%
  10.5555/186677.186689}, making the resulting reconstruction
algorithm decidable.
Moreover, since it is a fairly lightweight addition to the Hindley-Milner
algorithm, it composes well with other extensions.

One such extension is rank-N polymorphism, which has been observed by Xie
\emph{et al.}~\cite{DBLP:journals/pacmpl/XieCIL22} to be very useful in the
context of algebraic effects.
For example, consider a higher-order function that opens a file for
the duration of a computation received as its argument. A naive
approach could yield us the type
$\forall\tvarB\ldotp \texttt{Filepath} \to
  (\texttt{File} \to_{\texttt{IO}\effjoin\tvarB} \texttt{Unit})
  \to_{\texttt{IO}\effjoin\tvarB} \texttt{Unit}$. 
The effect $\texttt{IO}\effjoin\tvarB$ associated with the argument
and the entire function says that input/output effects can be
performed, as well as any effects described by the polymorphic
variable $\tvarB$. However, this type does not guarantee
that the file handle will not be used after it is closed. For example,
the argument could store the handle in some data structure for later
use.
Taking inspiration from Haskell's ST monad~\cite{DBLP:conf/haskell/WuSH14} and
the work of Xie \emph{et al.}, we can use rank-2 polymorphism to restrict the
scope of the handle, by rewriting the type as
$\forall\tvarB\ldotp \texttt{Filepath} \to
  (\ssforall{\tvarA}{
      \texttt{File}\;\tvarA \to_{\tvarA\effjoin\tvarB} \texttt{Unit}})
  \to_{\texttt{IO}\effjoin\tvarB} \texttt{Unit}$.
This time, the argument is polymorphic in the effect $\tvarA$, which
is associated with a particular file handle. Since all operations
on $\texttt{File}\;\tvarA$ have the effect $\tvarA$, they cannot
be used outside this function.

The considered example glosses over a certain difficulty with using
effect rows. We attempt to perform a concatenation of the variables
$\tvarA$ and $\tvarB$, but a row cannot contain two row variables.
One way of salvaging the ability to express this is to make $\alpha$
of a different kind (of atomic effects, which are elements of effect
rows). However, if we do that, the unification of rows stops being
unitary, somewhat compromising type and effect reconstruction.

The authors of the Flix programming language~\cite{DBLP:conf/onward/Madsen22}
chose a different approach.
Effects in Flix are set-like, with operations like union, intersection
and difference available as effect constructors.
Such a system is highly expressive, and admits a decidable inference algorithm
via boolean unification~\cite{DBLP:journals/pacmpl/MadsenP20}.
However, Flix does not currently support any form of higher-rank polymorphism.

Effect inference for set-like effects was thoroughly studied in the context of
static analysis~\cite{%
  DBLP:journals/iandc/TalpinJ94,%
  DBLP:conf/popl/TofteT94,%
  DBLP:conf/ccl/NielsonN94,%
  DBLP:conf/lomaps/NielsonNA96a,%
  DBLP:conf/sas/FahndrichA97,%
  DBLP:conf/fase/NielsonAN98}.
Notably, Talpin and Jouvelot~\cite{DBLP:journals/jfp/TalpinJ92} proposed an
algorithm that transforms the problem of effect inference into a corresponding
problem of satisfying subeffect constraints, which, for set-like effects,
behave like set inclusion.
Following Jouvelot and Gifford~\cite{DBLP:conf/popl/JouvelotG91}, their
algorithm implements a variant of ML-polymorphism with algebraic type schemes
of the form $\sforall{\tvarA_1\ldots\tvarA_n}{\constrs}{\type}$, which store
a set of constraints $\constrs$ together with universally quantified variables
$\tvarA_1\ldots\tvarA_n$.
This allows for solving constraints involving polymorphic variables
to be delayed to the point at which the scheme is instantiated.
Another interesting observation made in this area by Nielson and
Nielson~\cite{DBLP:books/daglib/0098888} is that subtyping induced by
subeffecting does not affect the shape of types.
They propose a two-stage approach to type-and-effect inference, where effects
are reconstructed once the type shapes are already known from the previous
phase.
Unfortunately, solutions introduced in this path of research were
designed for static analysis internal to the compiler, and not exposed
to the user directly. As a result, they had no need to support
rank-N polymorphism, which requires programmer-provided annotations.

In this paper, we present a sound and complete effect reconstruction algorithm
that 
needs minimal annotations to aid ergonomics,
supports higher-rank polymorphism for expressiveness,
and enjoys an intuitive set-like representation of effects.
Though none of the previously existing solutions fully satisfy these
requirements, we base our work on the results originating in the area of static
analysis.
Our algorithm is based on the work of Talpin and Jouvelot with its
algebraic type schemes, modified to use the two-stage approach.
The main difficulty arose from extending these techniques to support rank-N
polymorphism.

Widespread implementations of rank-N tend to require some annotations
to regain decidability of inference.
In those algorithms, whenever higher-rank polymorphism is required,
the annotations include, at the very least, both the polymorphic variable
\emph{binding}, as well as each \emph{bound occurrence}.
In contrast, our algorithm allows for rank-N effect polymorphism where only the
quantifiers have to be indicated, while the bound occurrences can be inferred.
This means that the annotation of the argument in the scheme can be
simplified from $\forall\tvarA\ldotp(\seffwildcard \to_\tvarA \seffwildcard)
  \to_{\tvarA} \seffwildcard$ to just $\forall\tvarA\ldotp \seffwildcard$.
As most programmers are not used to programming with effect systems at all,
requiring minimal amount of programmer-supplied effect information is fairly
important for encouraging greater adoption.

Designing an algorithm that supports the explosive mixture of subeffecting and
rank-N polymorphism with such minimal effect annotations is challenging due to
issues related to ubiquitous variable binding and a lack of the principal type
property in our setting.
As an example, suppose that we have a function \lstinline{f} known to be of
type
$(\texttt{Int} \to_{\texttt{IO}} \texttt{Int}) \to_{\texttt{DB}} \texttt{Int}$.
Now, consider the following function definition.
\begin{lstlisting}[mathescape]
let g (h : $\tforall{\tvarA} \texttt{Int} \to_{\seffwildcard} \texttt{Int}$) = h (f h)
\end{lstlisting}
Two possible different types for the function \lstinline{g} are
$(\tforall{\tvarA} \texttt{Int} \to_{\tvarA} \texttt{Int})
  \to_{\texttt{DB}}
  \texttt{Int}$
or
$(\tforall{\tvarA} \texttt{Int} \to_{\texttt{IO}} \texttt{Int})
  \to_{\texttt{IO}\effjoin\texttt{DB}}
  \texttt{Int}$,
but neither is more general than the other. Indeed, no principal type exists
for \lstinline{g} at all.

When further definitions using \lstinline{g} are added, it may turn out that
only one of these solutions is valid for the complete program.
Unfortunately,
as we are processing the annotation for \lstinline{h}, that information is not
available, and we do not even know whether the wildcard contains $\tvarA$ or not.
While algebraic type schemes provide a mechanism to delay the solving of some
constraints, it is not enough to delay decisions involving higher-rank polymorphic
variables, since they are confined to a smaller scope than the entire scheme of
a let-bound identifier.

In our solution to this problem we introduce two new mechanisms.
First, we introduce effect guards to delay the decision whether a locally-bound
variable should appear in a given effect. Second, when leaving the scope of
a quantified variable, we transform constraints involving this variable into
formulae, which do not contain effect variables, and can therefore be solved
later.

Our contributions can be summarized as follows.
\begin{itemize}
\item In \autoref{sec:full-problem} we propose a sound and complete algorithm
  for effect inference in the presence of higher-rank polymorphism.
  Our algorithm works even if we replace rank-N with the more general System~F
  polymorphism.
\item In \autoref{sec:no-constrs} we show that the problem of effect inference
  is decidable even when algebraic type schemes are replaced by simple type
  schemes without subeffect constraints.
\item As the considered problem is prone to subtle variable scoping issues,
  we have taken care to formalize all the presented results in the Rocq
  proof assistant.
  In \autoref{sec:formalization} we briefly describe
  the design decisions we made in our development.
\item We have successfully implemented the algorithm as a part of
  our \langname{}
  programming language---a realistic programming language
  with algebraic effects.
  In \autoref{sec:implementation}
  we summarize the insights that enable such a practical implementation.
\end{itemize}
The considered type system and proposed algorithm are quite large, so we
focus on the most important ideas in the main body of the paper. The full
definitions can be found in the appendices. The implementation of \langname{}
can be found at \url{https://github.com/fram-lang/dbl}.

\section{Simplified Setting}
\label{sec:simple}

Before we present a full algorithm,
let us start with a simpler problem of effect inference in a language
with ML-style effect polymorphism, but without any form of rank-N polymorphism.
In this section we present a variation of a well-established solution
based on the early work of Talpin and
Jouvelot~\cite{DBLP:journals/jfp/TalpinJ92}
that lays down the foundation
for our algorithm presented in the later sections.
The algorithm presented in this sections differs from the original formulation,
as it has been adjusted to better fit our setting and adapted to the two-stage
approach proposed by Nielson and Nielson~\cite{DBLP:books/daglib/0098888}.

\subsection{Syntax}

\begin{figure}[t!!]
  \begin{align*}
    \expr
        &\Coloneqq \varX 
        \mid \lam{\varX}{\stype}\expr
        \mid \expr\;\expr
        \mid \elet{\varX}{\expr}{\expr}
      \tag{expressions} \\
    \stype
        &\Coloneqq \tvarKT
        \mid \stype \tarrow{\seffect} \stype
      \tag{syntactic types} \\
    \seffect
        &\Coloneqq \tvarKE
        \mid \effpure
        \mid \seffect \effjoin \seffect 
        \mid \seffwildcard
      \tag{syntactic effects} \\
    \type
        &\Coloneqq \tvarKT
        \mid \type \tarrow{\effect} \type
      \tag{types} \\
    \effect
        &\Coloneqq \tvarKE
        \mid \effpure
        \mid \effect \effjoin \effect
      \tag{effects} \\
    \constrs &\Coloneqq \overline{\subeffect{\effect}{\effect}}
      \tag{sets of constraints} \\
    \scheme &\Coloneqq \sforall{\Delta^\keffect}{\constrs}\type
      \tag{type schemes}
  \end{align*}
  \caption{The syntax of the simplified calculus.}
  \label{fig:syntax}
\end{figure}

The syntax of the simplified calculus is presented in \autoref{fig:syntax}.
The calculus is a standard lambda calculus extended with the standard
polymorphic let-binding.
Lambda abstractions are annotated with \emph{syntactic types},
which are built from type variables ($\tvarKT$)
and arrows ($\stype_1 \tarrow{\seffect} \stype_2$)
annotated with \emph{syntactic effects} ($\seffect$),
which in turn are built from effect variables ($\tvarKE$),
the pure effect ($\effpure$), joins ($\seffect_1 \effjoin \seffect_2$),
and wildcards ($\seffwildcard$).
We use the same metavariables $\alpha$, $\beta$, $\gamma$ for
both type and effect variables, but we distinguish them by a kind annotation.
By convention, we omit the kind annotation when it is clear from the context.

Since syntactic effects contain wildcards,
we introduce separate syntactic categories of (internal) types and effects
used by the type system, with a similar grammar, except without wildcards.
Moreover, we consider two internal effects equal
if they can be proven equivalent
using the laws of an idempotent, commutative monoid with
join ($\effjoin$) as the monoidal operation, and the pure effect ($\effpure$)
as the neutral element. This guarantees a set-like semantics of effects.
The syntax of constraints and type schemes will be described later
in this section.

\subsection{Scopes and Substitutions}
\label{ssec:scopes}

We have found that in the presence of higher-rank polymorphism,
algorithms as well as their metatheories are
prone to errors related to variable binding, such as variable escape
or variable capture.
Therefore, in the technical presentation of this paper
as well as in the accompanying formalization
we are very precise about the scopes of variables.
To do so, we use the functorial approach to representing syntax~\cite{DBLP:conf/lics/FiorePT99}.
This is crucial for our Rocq formalization.
While in the paper we take more traditional approach,
some issues should be explained in the paper.

First, each syntactic category is in fact a family of sets of terms
indexed by sets of variables that are allowed to occur free.
For instance, the set of types is in fact a family of sets
$\{\types(A, B)\}_{A,B}$ indexed by
a set $A$ of possibly free effect variables,
and a set $B$ of possibly free type variables.
In order to avoid escaping variables,
each time when we write a term (\emph{e.g.,} a type),
there is an explicit or implicit object that ``binds''
these possibly free variables. For example, in case of a typing relation,
these variables are bound by the typing environment.

A \emph{substitution} is a function (or tuple of functions) from
\emph{all} potentially free variables to the substituted terms.
In the case of types, a substitution $\theta$ is a pair of functions
$\theta_e \colon A_1 \to \effects(A_2)$ and
$\theta_t \colon B_1 \to \types(A_2, B_2)$
that substitute for effect variables
and type variables respectively.
Note that the effects and types returned by $\theta_e$ and $\theta_t$
are themselves indexed by new sets of potentially free variables $A_2$ and
$B_2$, which need not be related to the domains of the functions, $A_1$ and
$B_1$, in any way.
We write $\bind{\theta} \type$ for
a capture-avoiding simultaneous substitution
for all free variables in $\type$ according to $\theta$.

When we use a term, for instance $\type \in \types(A, B)$, in a context
where more variables are available, \emph{e.g.,}
as an element of $\types(X \uplus A, B)$,
we have to apply a substitution $\iota_2\colon A \to X \uplus A$ that injects
the old variables into a larger set.
While in the formalization we keep this precise,
in the paper we omit obvious projections, injections, and permutations
to reduce noise.
Similarly, we omit identity-like parts of substitutions
whenever we substitute only for a subset of the available variables.
For instance, when substituting $\effect$
for $\alpha$ in $\type \in \types(\{\alpha\}\uplus A, B)$, we write
$\Subst{\alpha}{\effect}{\type}$, which resembles the standard notation
for substitution\footnote{
  We decided to use this
  bind-like notation ($\Subst{\alpha}{\effect}{\type}$)
  instead of the standard notation ($\type \{\alpha \mapsto \effect\}$)
  in order to keep the notation consistent.
}.

\subsection{Declarative Type System}

Here we present a type and effect system, which serves as a specification
for the effect inference algorithm presented later in this section.
We call this system \emph{declarative}, because it doesn't need to be
algorithmic (we allow non-syntax directed rules, and guessing data which
are not part of the input), but it should be simple and intuitive.
We start from some auxiliary relations.
Due to space constraints, we omit most of the obvious rules,
and focus on presenting the main ideas and the most notable rules.
We refer the reader to Appendix~\ref{appendix:system}
for the full definition of the system.

\paragraph{Type and effect matching.}

\begin{figure}[t!!]
  \reldefline{Effect Matching}
    {$\EnvRelMatch{\Delta}{\seffect}{\effect}$}
  \begin{mathpar}
    \inferrule{ }{\EnvRelMatch{\Delta}{\tvarKE}{\tvarKE}}

    \inferrule{ }{\EnvRelMatch{\Delta}{\effpure}{\effpure}}

    \inferrule{
      \EnvRelMatch{\Delta}{\seffect_1}{\effect_1}
      \and
      \EnvRelMatch{\Delta}{\seffect_2}{\effect_2}}
      {\EnvRelMatch{\Delta}{\seffect_1 \effjoin \seffect_2}
        {\effect_1 \effjoin \effect_2}}

    \inferrule{ }{\EnvRelMatch{\Delta}{\seffwildcard}{\effect}}
  \end{mathpar}
  \reldefline{Type Matching}
    {$\EnvRelMatch{\Delta}{\stype}{\type}$}
  \begin{mathpar}
    \inferrule{ }{\EnvRelMatch{\Delta}{\tvarKT}{\tvarKT}}

    \inferrule{
      \EnvRelMatch{\Delta}{\stype_1}{\type_1}
      \and
      \EnvRelMatch{\Delta}{\stype_2}{\type_2}
      \and
      \EnvRelMatch{\Delta}{\seffect}{\effect}}
      {\EnvRelMatch{\Delta}{\stype_1 \tarrow{\seffect} \stype_2\ }
        {\ \type_1 \tarrow{\effect} \type_2}}
  \end{mathpar}
  \caption{Type and effect matching.}
  \label{fig:matching}
\end{figure}

A programmer may omit some effect annotations using wildcard syntactic
effects. The declarative system ``guesses'' the internal effects
that should appear in place of these wildcards.
We accomplish this with the \emph{effect matching} relation
$\EnvRelMatch{\Delta}{\seffect}{\effect}$
defined in \autoref{fig:matching}.
The relation uses a type environment ($\Delta$), which keeps track
of the available type and effect variables in the current scope,
and says that the syntactic effect~$\seffect$ matches the internal
effect~$\effect$.
As mentioned in \autoref{ssec:scopes}, we implicitly assume
that both syntactic and internal effects are well-formed in $\Delta$.
For instance, in the first rule we implicitly assume that $\tvarKE\in\Delta$,
but we omit such premises to avoid clutter.

The definition of effect matching consists of three obvious structural
rules, and the last rule saying that a wildcard can be matched by
any (well-formed) effect.
On top of this relation we define \emph{type matching}.
Since we assume that the types were inferred in the previous phase,
the definition of this relation consists of structural rules only.

\paragraph{Constraints, subeffecting, and subtyping.}
The key feature of the presented type system is that it allows for
abstracting over subeffecting constraints
of the form $\subeffect{\effect_1}{\effect_2}$.
Therefore, the \emph{subeffecting relation} (with the judgement of the form
\mbox{$\EnvRelSubEffect{\Delta}{\constrs}{\effect_1}{\effect_2}$})
uses another environment $\constrs$ which is a set of abstracted
subeffecting constraints.
For a fixed $\Delta$ and $\constrs$,
the subeffecting relation
$\EnvRelSubEffect{\Delta}{\constrs}{\effect_1}{\effect_2}$
is defined as the least preorder on effects
containing $\constrs$, with $\effjoin$ as the semilattice join,
and $\effpure$ as the least element. The full definition can be found
in the Appendix.
Using this relation we can also define the constraint set entailment relation
$\EnvRel{\Delta}{\constrs}{\constrs'}$ by ensuring that subeffecting
holds for every constraint in $\constrs'$.
Moreover, subeffecting induces subtyping in the standard way:
the definition of $\EnvRelSubType{\Delta}{\constrs}{\type_1}{\type_2}$
consists of the standard structural rules with contravariant
left-hand-side of the arrow type.

\paragraph{Type schemes.}

As observed by Jouvelot and Gifford,
a reasonable effect reconstruction algorithm may be unable to
solve generated constraints while examining a
let binding~\cite{DBLP:conf/popl/JouvelotG91}.
To address this problem, they propose \emph{algebraic type schemes}
to be attached to let-bound variables.
In their approach a type scheme abstracts constraints
that cannot be solved yet, and requires that these constraints are
satisfied at the place of the scheme instantiation.

As the last ingredient of the syntax presented in \autoref{fig:syntax}
our type schemes follow this path.
A type scheme $\sforall{\Delta^\keffect}{\constrs}{\type}$
abstracts a set of effect variables $\Delta^\keffect$
and remembers a set of subeffecting constraints $\constrs$
that may use variables bound in $\Delta^\keffect$.
We write just $\tau$ when both $\Delta^\keffect$ and $\constrs$
are empty.
Since we assume a two-stage approach to type-and-effect inference,
type schemes do not bind type variables:
type polymorphism can be handled in the previous phase,
and expressed using rank-N polymorphism introduced in
\autoref{sec:full-problem}.

\paragraph{Typing relation.}

\begin{figure}[t!!]
  \reldefline{Typing}
    {$\TypingRel{\Delta}{\constrs}{\Gamma}{\expr}{\type}{\effect}$}
  \begin{mathpar}
    \inferrule{
      \Gamma(\varX)=\sforall{\Delta'}{\constrs'}\type
      \and
      \Delta;\constrs\vdash\bind{\theta}\constrs'}
      {\TypingRel{\Delta}{\constrs}{\Gamma}
        {\varX}{\bind{\theta}\type}{\effpure}}

    \inferrule{
      \EnvRelMatch{\Delta}{\stype}{\type_1}
      \and
      \TypingRel{\Delta}{\constrs}{\Gamma,\varX:\type_1}
          {\expr}{\type_2}{\effect}}
      {\TypingRel{\Delta}{\constrs}{\Gamma}{\lam{\varX}{\stype}\expr}
        {\type_1\to_{\effect}\type_2}{\effpure}}

    \inferrule{
      \TypingRel{\Delta}{\constrs}{\Gamma}{\expr_1}
          {\type_2\to_{\effect}\type_1}{\effect}
      \and
      \TypingRel{\Delta}{\constrs}{\Gamma}{\expr_2}
          {\type_2}{\effect}}
      {\TypingRel{\Delta}{\constrs}{\Gamma}{\expr_1\;\expr_2}
        {\type_1}{\effect}}

    \inferrule{
      \TypingRel{\Delta,\Delta^\keffect_g}{\constrs,\constrs_g}
          {\Gamma}{\expr_1}{\type_1}{\effpure}
      \\\\
      \TypingRel{\Delta}{\constrs}
        {\Gamma, \varX:\sforall{\Delta^\keffect_g}{\constrs_g}{\type_1}}{\expr_2}
          {\type_2}{\effect}}
      {\TypingRel{\Delta}{\constrs}{\Gamma}
        {\elet{\varX}{\expr_1}{\expr_2}}{\type_2}{\effect}}

    \inferrule{
      \TypingRel{\Delta}{\constrs}{\Gamma}{\expr}{\type}{\effect} \\\\
        \EnvRelSubType{\Delta}{\constrs}{\type}{\type'} \\\\
        \EnvRelSubEffect{\Delta}{\constrs}{\effect}{\effect'}}
      {\TypingRel{\Delta}{\constrs}{\Gamma}{\expr}{\type'}{\effect'}}
  \end{mathpar}
  \caption{Typing relation.}
  \label{fig:typing}
\end{figure}

The definition of the typing relation is given in \autoref{fig:typing}.
In the variable rule, the scheme assigned to a variable
is immediately instantiated with any substitution $\theta$
that satisfies the constraints attached to the scheme.
In the $\lambda$-abstraction rule, the variable is annotated with
a syntactic type~$\stype$,
so the actual (monomorphic) type~$\type_1$ assigned to the variable
must match the annotation.
As usual for type-and-effect systems,
the $\lambda$-abstraction is pure, but the effect of the body is
remembered in the arrow type.
The application rule is standard.
The let-rule resembles the one known from ML-style polymorphism,
but deserves additional explanation.
First, it allows to implicitly abstract over some effect variables
$\Delta^\keffect_g$ and constraints $\Omega_g$ in the first expression.
Moreover, while it allows polymorphic definitions,
it enforces \emph{purity restriction}~\cite{DBLP:conf/aplas/AsaiK07}
in order to ensure that the type system is sound.
The last rule is the standard subsumption rule,
which allows to change the type and effect of an expression
to a supertype and supereffect.

It can be shown that the presented type system is sound wrt
the standard operational semantics (not presented here) using
the standard methods of logical relations~\cite{DBLP:conf/popl/AppelMRV07,%
DBLP:journals/pacmpl/BiernackiPPS18}
or progress and preservation~\cite{DBLP:journals/iandc/WrightF94,%
DBLP:conf/popl/Leijen17}.
However, these results lie outside the scope of this paper,
and therefore we focus our attention on the algorithmic effect inference.

\subsection{Algorithmic Effect Inference}

The effect inference algorithm proposed by 
Talpin and Jouvelot~\cite{DBLP:journals/jfp/TalpinJ92}
is a modification of the well-known
Hindley-Milner algorithm~$\mathcal{W}$~\cite{DBLP:journals/jcss/Milner78}.
In the algorithm~$\mathcal{W}$ unknown types are represented by unification
variables, which may be instantiated by the unification procedure.
Similarly, the presented algorithm uses unification variables to represent
unknown effects, and such variables may be instantiated by
solving subeffecting constraints.
While for convenience in our implementation
we use a separate syntactic category for unification variables,
in the presentation in this paper, as well as in the formalization,
we use regular effect variables for that purpose,
but we explicitly keep track which variables were generated.

For being more precise, when the algorithm examines some term
defined over the set of effect variables $\Delta$, it returns a
tuple \lstinline[mathescape=true]{($\Delta_g$; $\ldots$)}
whose other elements are defined over the set of effect variables
$\Delta \uplus \Delta_g$.
Let us start with some auxiliary functions, which correspond
to the auxiliary relations of the declarative system.

\paragraph{Type and effect matching.}

\begin{figure}[t!!]
\begin{minipage}{.45\textwidth}
\begin{lstlisting}[language=meta,mathescape=true]
tr_effect($\tvarA$) = ($\varnothing$; $\tvarA$)
tr_effect($\effpure$) = ($\varnothing$; $\effpure$)
tr_effect($\seffwildcard$) =
  fresh $\alpha$ in ($\{\alpha\}$; $\alpha$)
\end{lstlisting}
\end{minipage}
\begin{minipage}{.45\textwidth}
\begin{lstlisting}[language=meta,mathescape=true]
tr_effect($E_1 \effjoin E_2$) =
  let ($\Delta_1$; $\effect_1$) = tr_effect($E_1$) in
  let ($\Delta_2$; $\effect_2$) = tr_effect($E_2$) in
  ($\Delta_1 \uplus \Delta_2$; $\effect_1 \effjoin \effect_2$)
\end{lstlisting}
\end{minipage}
  \caption{Translation of syntactic effects.}
  \label{fig:tr-effect}
\end{figure}

Effect matching is realized by the function that translates
syntactic effect to the internal representation
and is given in \autoref{fig:tr-effect}.
The function is defined by structural recursion on syntactic effects,
and returns a tuple containing generated effect variables and translated
effects.
Most cases are straightforward and just replace each syntactic
effect construct with its corresponding internal effect construct.
The only interesting case is for wildcards, where the function
``guesses'' the effect that should be returned. This ``guessing''
is realized by generating a fresh variable,
similarly to how type guessing is implemented
in the original algorithm~$\mathcal{W}$.

Using effect matching, we define the function \lstinline{tr_type},
which translates types. The definition consists of structural cases only,
so we omit it in our presentation.
Precise definition of this, as well as other functions can be found
in Appendix~\ref{appendix:algo}.

\paragraph{Constraints, subeffecting, and subtyping.}

Subtyping is realized by a function \lstinline{subtype} that collects
subeffecting constraints that need to be satisfied in order to
make the relevant subtyping judgment hold.
For arrow types this function is defined as follows.
\begin{lstlisting}[language=meta,mathescape=true]
subtype($\type_1' \tarrow{\effect_1} \type_1
    $; $\type_2' \tarrow{\effect_2} \type_2$) =
  let $\Omega_1$ = subtype($\type_2'$; $\type_1'$) in
  let $\Omega_2$ = subtype($\type_1$; $\type_2$) in
  $\Omega_1 \cup \Omega_2 \cup \{ \subeffect{\effect_1}{\effect_2} \}$
\end{lstlisting}
Since there are no effect binders in types,
the other cases are straightforward and omitted in the presentation.

\paragraph{The algorithm.}

\begin{figure}[t!!]
\begin{lstlisting}[language=meta,mathescape=true]
infer($\Gamma$; $\varX$) =
  let ($\sforall{\Delta_1}{\Omega_1} \type$) = $\Gamma(\varX)$ in
  ($\Delta_1$; $\type$; $\effpure$; $\Omega_1$)

infer($\Gamma$; $\lam{\varX}{\stype} e$) =
  let ($\Delta_1$; $\type_1$) = tr_type($\stype$) in
  let ($\Delta_2$; $\type_2$; $\effect$; $\Omega
    $) = infer($\Gamma, x : \type_1$; $e$) in
  ($\Delta_1 \uplus \Delta_2$; $
    \type_1 \tarrow{\effect} \type_2$; $\effpure$; $\Omega$)

infer($\Gamma$; $e_1\;e_2$) =
  let ($\Delta_1$; $\type_1$; $\effect_1$; $\Omega_1
    $) = infer($\Gamma$; $e_1$) in
  match $\type_1$ with
  | $\type_a \tarrow{\effect} \type_v$ =>
    let ($\Delta_2$; $\type_2$; $\effect_2$; $\Omega_2
      $) = infer($\Gamma$; $e_2$) in
    let $\Omega_s$ = subtype($\type_2$; $\type_a$) in
    ($\Delta_1 \uplus \Delta_2$; $
      \type_v$; $\effect_1 \effjoin \effect_2 \effjoin \effect$; $
      \Omega_1 \cup \Omega_2 \cup \Omega_s$)
  | _ => fail

infer($\Gamma$; {let} $\varX$ = $e_1$ {in} $e_2$) =
  let ($\Delta_1$; $\type_1$; $\effect_1$; $\Omega_1
    $) = infer($\Gamma$; $e_1$) in
  infer($\Gamma, \varX : \sforall{\Delta_1}{
    \Omega_1 \cup \{\subeffect{\effect_1}{\effpure}\}}
    \type_1$; $e_2$)
\end{lstlisting}
  \caption{The effect inference algorithm for the simplified problem.}
  \label{fig:simple-algo}
\end{figure}

Now we are in position to present the sound and complete effect inference
algorithm for the simplified problem of this section.
The algorithm is presented in \autoref{fig:simple-algo}.
The function \lstinline{infer}
takes the typing environment and an expression
and returns tuple of the form
\lstinline[mathescape=true]{($\Delta$; $\type$; $\effect$; $\Omega$)}
containing set of generated variables that may appear in other
components of the tuple: inferred type, effect,
and the set of constraints that should be satisfied to make
typing judgement hold.
The function is defined by a structural recursion on the expression.

In the variable case, its scheme ($\sforall{\Delta_1}{\Omega_1}{\type}$)
is instantiated with fresh effect variables.
Since variables introduced by opening a binder are always fresh by convention,
it is enough to return the set $\Delta_1$ as the generated variables,
together with the type~$\type$ and the pure effect.
In the declarative system, the constraints~$\Omega_1$
needs to be satisfied, so the algorithm includes $\Omega_1$ in the
returned~tuple.

The type annotation for the $\lambda$-abstraction argument needs to be translated
to the internal representation and this process may generate new
variables ($\Delta_1$).
The recursive call to the body is performed in the context
where these variables are available and may also generate new variables
($\Delta_2$).
Both sets $\Delta_1$ and~$\Delta_2$ are combined in the final result.
The case for function application is standard and
as usual for systems with subtyping, allows expressions of smaller type
in the argument position.
Note that thanks to two-stage approach,
subtype-checking generates only subeffecting constraints.
These constraints can be propagated in the result tuple,
instead of solving them immediately.

Now we focus on the let-binding case that introduces effect-polymorphism
into the language. For the simplified calculus it is surprisingly simple:
all variables and constraints generated during
the first recursive call are abstracted and put into
the algebraic type scheme.
The constraint set in the scheme is extended with
$\subeffect{\effect_1}{\effpure}$
in order to ensure that expressions $\expr_1$ is pure.

\begin{remark}
  In the presented algorithm, we generalize all the variables generated
  during type-checking of the body of the let binding.
  This differs from the original presentation of Talpin-Jouvelot algorithm
  as well as classical Hindley-Milner algorithm,
  where only the variables that do not escape through the environment
  are generalized.
  We can do so, because (a) our algorithm doesn't return any substitution
  and (b) all generated constraints can be included in the type scheme,
  so generated variables have no mean to escape through the environment.
  In the algorithm presented in the next section, the condition~(b)
  will be no longer true, so additional step of constraint factorization
  will be required.
  Interestingly, the classical Hindley-Milner algorithm
  can be presented in the style of the algorithm from this section
  with additional step of factorizing substitution into its kernel
  and additional renaming. However, the details are outside the scope
  of this paper.
\end{remark}

The presented algorithm is sound and complete with respect to the
declarative type system, which can be stated formally by the following
two theorems.

\begin{theorem}[Soundness]
  For the expression $e$ defined over the set of type variables~$\Delta$,
  if $\codeinfer{\Gamma}{e} = \codetupIV{\Delta'}{\type}{\effect}{\Omega}$,
  then
  $\TypingRel{\Delta,\Delta'}{\Omega}{\Gamma}{e}{\type}{\effect}$.
\end{theorem}

\begin{theorem}[Completeness]
  If $\TypingRel{\Delta}{\Omega}{\Gamma}{e}{\type}{\effect}$,
  then $\codeinfer{\Gamma}{e} =
  \codetupIV{\Delta'}{\type'}{\effect'}{\Omega'}$
  for some $\Delta'$, $\type'$, $\effect'$, and $\Omega'$.
  Moreover, there exists substitution~$\theta$
  that maps effect variables from $\Delta'$ to effects over $\Delta$,
  such that
    $\EnvRel{\Delta}{\Omega}\bind{\theta}\constrs'$,
    $\EnvRelSubType{\Delta}{\Omega}{\bind{\theta}\type'}{\type}$,
    and $\EnvRelSubEffect{\Delta}{\Omega}{\bind{\theta}\effect'}{\effect}$.
\end{theorem}

\section{Effect Inference with Higher-Rank Polymorphism}
\label{sec:full-problem}

A major increase in difficulty arises when we introduce two changes
into the underlying type system: higher-rank polymorphism
announced in the introduction,
and restriction of constraints that can be included in type schemes.
Both changes are motivated by practical reasons.

\paragraph{Higher-rank polymorphism.}

Usually, practical type systems with type reconstruction restrict
the expressiveness of parametric polymorphism.
For instance, the Hindley-Milner type system allows only
polymorphic types in the prenex form, and its more liberal extension
to rank-N, distinguish between monomorphic types and polymorphic type schemes,
and allows polymorphic variables
to be instantiated with monomorphic types only.
Since we use two-stage approach, we assume that the types are already
inferred, so decisions on particular restrictions are left to the designer
of the underlying type system (without effects).
Such decisions seem to be orthogonal to the problem of effect inference,
so for simplicity we extend the system with a general form of polymorphism
in the style of System~F.
We extend the grammar of expressions, syntactic types, and types as
follows.
\begin{align*}
  \expr
      &\Coloneqq \ldots
      \mid \lamT{\alpha}\expr
      \mid \lamE{\alpha}\expr
      \mid \expr\appT{\stype}
      \mid \expr\appE{\seffect}
    \tag{expressions} \\
  \stype
      &\Coloneqq \ldots
      \mid \tforallT{\alpha}\stype
      \mid \tforallE{\alpha}\stype
    \tag{syntactic types} \\
  \type
      &\Coloneqq \ldots
      \mid \tforallT{\alpha}\type
      \mid \tforallE{\alpha}\type
    \tag{types}
\end{align*}
Types and syntactic types are extended with polymorphic quantifier,
that can abstract both type and effect variables.
Places where such polymorphic type is introduced or eliminated are
explicitly marked in the syntax of expressions
by a type or effect $\lambda$-abstraction and application.
The type application ($e\appT{\stype}$) contains
syntactic type, as it contains complete information
inferred by a previous type-inference phase and additionally may
contain effects provided by the user.
Similarly, in the effect application ($e\appE{\seffect}$), the syntactic
effect may be just a wildcard,
but it can be also more precise.

\begin{figure}[t!!]
  \reldefline{Type Matching}
    {$\EnvRelMatch{\Delta}{\stype}{\type}$}
  \begin{mathpar}
    \inferrule{
      \EnvRelMatch{\Delta, \alpha^\ktype}{\stype}{\type}}
      {\EnvRelMatch{\Delta}{\tforallT{\alpha} \stype}{\tforallT{\alpha} \type}}

    \inferrule{
      \EnvRelMatch{\Delta, \alpha^\keffect}{\stype}{\type}}
      {\EnvRelMatch{\Delta}{\tforallE{\alpha} \stype}{\tforallE{\alpha} \type}}
  \end{mathpar}
  \reldefline{Subtyping}
    {$\EnvRelSubType{\Delta}{\Omega}{\type}{\type}$}
  \begin{mathpar}
    \inferrule{
      \EnvRelSubType{\Delta, \alpha^\ktype}{\constrs}{\type_1}{\type_2}}
      {\EnvRelSubType{\Delta}{\constrs}
        {\tforallT{\alpha}{\type_1}}
        {\tforallT{\alpha}{\type_2}}}

    \inferrule{
      \EnvRelSubType{\Delta, \alpha^\keffect}{\constrs}{\type_1}{\type_2}}
      {\EnvRelSubType{\Delta}{\constrs}
        {\tforallE{\alpha}{\type_1}}
        {\tforallE{\alpha}{\type_2}}}
  \end{mathpar}
  \reldefline{Typing}
    {$\TypingRel{\Delta}{\Gamma}{\constrs}{\expr}{\type}{\effect}$}
  \begin{mathpar}
    \inferrule{
      \TypingRel{\Delta,\tvarA^\ktype}{\constrs}
        {\Gamma}{\expr}{\type}{\effpure}}
      {\TypingRel{\Delta}{\constrs}{\Gamma}{\lamT{\tvarA}\expr}
        {\tforallT{\tvarA}\type}{\effpure}}

    \inferrule{
      \TypingRel{\Delta,\tvarA^\keffect}{\constrs}
        {\Gamma}{\expr}{\type}{\effpure}}
      {\TypingRel{\Delta}{\constrs}{\Gamma}{\lamE{\tvarA}\expr}
        {\tforallE{\tvarA}\type}{\effpure}}

    \inferrule{
      \TypingRel{\Delta}{\constrs}{\Gamma}{\expr}
          {\tforallT{\tvarA}{\type}}{\effect}
      \and
      \EnvRelMatch{\Delta}{\stype}{\type'}}
      {\TypingRel{\Delta}{\constrs}{\Gamma}{\expr \appT{\stype}}
        {\Subst{\tvarA}{\type'}{\type}}{\effect}}

    \inferrule{
      \TypingRel{\Delta}{\constrs}{\Gamma}{\expr}
          {\tforallE{\tvarA}{\type}}{\effect}
      \and
        \EnvRelMatch{\Delta}{\seffect}{\effect'}}
      {\TypingRel{\Delta}{\constrs}{\Gamma}{\expr \appE{\seffect}}
        {\Subst{\tvarA}{\effect'}{\type}}{\effect}}
  \end{mathpar}
  \caption{New inference rules for explicit polymorphism.}
  \label{fig:poly-rules}
\end{figure}

The new constructs come with new inference
rules presented in \autoref{fig:poly-rules}.
The new matching and subtyping rules are structural,
but since they extend the type environment ($\Delta$) in the premise,
their presence has serious consequence:
wildcards under the quantifier $\tforallE{\alpha}{\stype}$
can be matched with effects containing~$\alpha^\keffect$.
The new typing rules are mostly standard, with a minor exception
that in the application rules the matching relation is used
in order to translate syntactic representation of type or effect
to the internal one.

\paragraph{Type schemes revisited.}

The algebraic type schemes ($\sforall{\Delta}{\Omega}\type$)
from the previous section
allowed abstracting any set of constraints $\Omega$.
While this decision led to a simple, sound, and complete effect inference
algorithm, the obtained system is impractical and counter-intuitive.
The main reason is that the type scheme may abstract contradictory
constraints, \emph{e.g.,} $\subeffect{\texttt{IO}}{\effpure}$,
delaying reporting programming errors to unexpected moments.
As an extreme case, a mistake in a library function
may be reported to the user of the library, but not to the developer.

In order to avoid such strange behaviour,
we should restrict somehow the constraints
that can be included in a type scheme.
Such a restriction should (a)~disallow contradictory constraints
and (b)~be intuitive to the programmer.
One could say that the Holy Grail would be disallowing constraints at all,
and using type scheme of the form $\ssforall{\Delta}{\type}$
like in the classical Hindley-Milner type system.
In \autoref{sec:no-constrs} we show that with such a restriction
we can still obtain a sound and complete algorithm,
but the price we pay seems to be unacceptable.

In this section we propose another restriction.
We allow only constraints in the form of upper-bounds of
effect variables bound by the scheme.
Formally, the scope-aware definition of valid type scheme is the following.
\begin{align*}
  \ConstrUB{\Delta}{\Delta'} & \eqdef
    \{ \subeffect{\alpha}{\effect} \mid \alpha \in \Delta'
      \wedge \effect \in \effects(\Delta \uplus \Delta')\} \\
  \ValidScheme{\Delta}      & \eqdef
    \{ \sforall{\Delta'}{\Omega}{\type} \mid
      \forall (\subeffect{\effect_1}{\effect_2}) \in \Omega\ldotp
        (\subeffect{\effect_1}{\effect_2}) \in \ConstrUB{\Delta}{\Delta'} \}
\end{align*}
For the purpose of this section we assume that the type schemes
assigned to let-bound variables are valid.

Upper-bound constraints are never contradictory,
because the are trivially satisfied, when we substitute pure
effects for variables bound by the scheme.
Moreover, type schemes with upper-bound constraints can give
a quite precise information about function behavior,
as in the following example.

\begin{example}
  Assume we have function \lstinline[mathescape=true]{callLater : (Unit $
    \tarrow{\texttt{IO}}$ Unit) $\tarrow{\texttt{DB}}$ Unit}
  that registers given function is some data structure (effect \lstinline{DB})
  for calling it later, where the effect \lstinline{IO} is available.
  Consider the following function.
\begin{lstlisting}
let callNowOrLater now f =
  if now then f ()
  else callLater f
\end{lstlisting}
  It can accept the function with the \lstinline{IO} effect,
  but there is nothing wrong in passing the pure function.
  In the latter case, the call to \lstinline{callNowOrLater} would
  not perform \lstinline{IO} effect, so there is no need to pollute
  its effect with \lstinline{IO}.
  It can be achieved by assigning the type scheme
  \[
    \sforall{\alpha}{\subeffect{\alpha}{\texttt{IO}}}
      \texttt{Bool} \to (\texttt{Unit} \tarrow{\alpha} \texttt{Unit})
        \tarrow{\alpha \effjoin \texttt{DB}} \texttt{Unit}\textrm{.}
  \]
  Such a general type scheme cannot be expressed
  in the type system without constraints.
\end{example}

\subsection{Effect Guards}

Before presenting the algorithm we will try to grab some intuitions
that will help us understand the problem we face.
Let us go back to one of motivating examples, when the user provided
a syntactic type annotation
$\stype \eqdef \tforallE{\alpha}\beta \tarrow{\seffwildcard} \beta$.
While translating it to the internal representation, a reasonable algorithm
would generate a fresh variable $\gamma$ in place of
the wildcard.
However, problems arise when it comes to leave the scope of the quantifier.
The syntactic type $\stype$ is matched by
both $\tforallE{\alpha}\beta \tarrow{\gamma} \beta$
and $\tforallE{\alpha}\beta \tarrow{\gamma \effjoin \alpha} \beta$,
but neither of them is more general than another.
The algorithm should decide if it should include $\alpha$ in the effect
matched with the wildcard.
The idea behind our algorithm is to delay such decisions
by conditionally including bound variable in places where it can appear.
To do so, we extend the grammar of effects used by the algorithm
by the construct $\effect\effguard\form$ called \emph{effect guard},
where $\form$ is a formula of the propositional logic.
\begin{align*}
  \form
      &\Coloneqq \fvarP \mid \ftop \mid \bot
      \mid \form\land\form
      \mid \form\lor\form
      \mid \form\Rightarrow\form
      \tag{formulae} \\
  \effect
      &\Coloneqq \ldots
      \mid \effect\effguard\form
      \tag{effects}
\end{align*}
The grammar of effects used by the declarative type system remain unchanged.
Intuitively, the effect $\effect\effguard\form$ means $\effect$
when the formula $\form$ is satisfied, and $\effpure$ otherwise.
With this new construct, the most general type that can be matched
with $\stype$ would be
$\tforallE{\alpha}
  \beta \tarrow{\gamma \effjoin (\alpha \effguard \fvarP)} \beta$,
where $\fvarP$ is a fresh propositional variable.

As we extend the syntax of effects, for the purpose of the metatheory
of the algorithm we define a version of a declarative type system
that take into account the extended grammar of effects.
For a given valuation $\rho$ of propositional variables
we define matching, subeffecting, subtyping, and typing relations
(all denoted with $\vdash_{\!\rho}$) all defined by the rules analogous
to the original declarative system, with the following three
subeffecting rules that handle additional effect guard construct.
\begin{mathpar}
  \inferrule
    {\rho\not\models\form}
    {\RhoRelSubEffect{\Delta}{\Omega}{\effect_1\effguard\form}{\effect_2}} 

  \inferrule
    {\RhoRelSubEffect{\Delta}{\Omega}{\effect_1}{\effect_2}}
    {\RhoRelSubEffect{\Delta}{\Omega}{\effect_1\effguard\form}{\effect_2}}

  \inferrule
    { \rho\models\form \and
      \RhoRelSubEffect{\Delta}{\Omega}{\effect_1}{\effect_2}}
    {\RhoRelSubEffect{\Delta}{\Omega}{\effect_1}{\effect_2\effguard\form}}
\end{mathpar}
The first rule says that guarded effect is pure, when the formula is false.
The second and the third rule say that guard can be
discarded if the formula is true.
Note that in the second rule, the there is no premise $\rho\models\form$
for simplicity: the conclusion always holds when the formula is false,
by the first rule.

In order to motivate the second ingredient of our solution,
let us consider when the
type $\tforallE{\alpha} \beta \tarrow{\alpha \effguard \form_1} \beta$
is a subtype of
$\tforallE{\alpha}\beta \tarrow{\alpha \effguard \form_2} \beta$.
The algorithm from \autoref{sec:simple} simply collects subeffecting
constraints that are required the subtyping to hold.
In this case we cannot simply return the constraint
$\subeffect{\alpha \effguard \form_1}{\alpha \effguard \form_2}$,
because $\alpha$ would escape its scope.
However, we can observe that this constraint is satisfied for each effect
substituted for $\alpha$ if and only
if the implication $\form_1 \Rightarrow \form_2$ is true.
We can see this implication as another form of constraint,
which does not contain effect variables, and can be propagated outside
the quantifier.

\begin{figure}[t!!]
  \begin{equation*}
    \begin{aligned}
      (\tvarKE)\toFormula{\tvarKE} & \eqdef \top \\
      (\effpure)\toFormula{\tvarKE} & \eqdef \bot \\
      (\effect\effguard\form)\toFormula{\tvarKE] & \eqdef \effect[\tvarKE}\land\form
    \end{aligned}\qquad
    \begin{aligned}
      (\tvarB^{\keffect})\toFormula{\tvarKE}
        & \eqdef \bot \quad (\tvarKE\neq\tvarB^{\keffect}) \\
      (\effect_1\effjoin\effect_2)\toFormula{\tvarKE}
        & \eqdef \effect_1\toFormula{\tvarKE]\lor\effect_2[\tvarKE} \\
      \constrs\toFormula{\tvarKE} &\eqdef
      \bigwedge_{\quad\mathclap{(\subeffect{\effect_1}{\effect_2})\in\constrs}\quad}
          \left(
          \effect_1\toFormula{\tvarKE}\Rightarrow\effect_2\toFormula{\tvarKE}
          \right)
    \end{aligned}
  \end{equation*}
  \caption{Extracting formulae from effects.}
  \label{fig:to-formula}
\end{figure}

For a systematic method of generating formula when leaving the scope
of the effect variable, we define the operation of extracting
formulae guarding this variable in an effect.
The definition is given in \autoref{fig:to-formula}.
It simply builds the formula that is true when the effect variable
is present in the effect.
We extend this definition to (finite) sets of constraints,
by taking a conjunction of implications.

\subsection{Algorithmic Effect Inference}

Once we have intuitive understanding of required tools
we proceed to presenting the algorithm.
In the presentation we focus mostly on the effect polymorphism,
because---as we have seen above---scope leaving requires special attention.
On the other hand, type polymorphism doesn't pose similar problems,
so we omit it in the presentation.
However, we refer interesting reader to the technical appendix
for the details of the whole construction.

\paragraph{Type matching.}

As we observed, the key idea of our algorithm is to delay some decisions
by introducing predicates of propositional logic.
Therefore, some functions, like \lstinline{tr_type}
generate fresh propositional variables when dealing with problematic
effect binders.
Similarly to effect variables, we explicitly keep track
which propositional variables were generated by the algorithm,
so now, the \lstinline{tr_type} function returns triples
of the form \lstinline[mathescape=true]{($P$; $\Delta$; $\type$)},
where $P$ is the set of generated propositional variables,
while the meaning of other components remains unchanged.
For most cases, the set of generated propositional variables
is a union of sets generated by subexpressions (and empty when there
are no subexpressions).
The only interesting case is the following case for
effect-polymorphic types.
\begin{lstlisting}[language=meta,mathescape=true]
tr_type($\tforallE{\alpha} \stype$) =
  let ($P$; $\Delta$; $\type$) = tr_type($\stype$) in
  fresh $P_s$ = $\{ \fvarP_\beta \mid \beta \in \Delta \}$ in
  fresh $\Delta'$ = $\{ \gamma_\beta \mid \beta \in \Delta \}$ in
  ($P \uplus P_s$; $\Delta'$; $\tforallE{\alpha} \bind{\subst{
    \overline{\beta \in \Delta \mapsto \gamma_\beta
      \effjoin \alpha \effguard \fvarP_\beta}}} \type $)
\end{lstlisting}

When leaving the scope of variable $\alpha$, the algorithm should decide
for each effect matched by wildcard if it should contain $\alpha$.
Such effects are represented by effect variables from $\Delta$,
so the algorithm delay the decision, by extending each effect containing
$\beta\in\Delta$ by $\alpha\effguard \fvarP$. The fresh propositional variable
$\fvarP$ represents the decision: if it is true,
$\alpha$ is included in the effect, otherwise $\alpha \effguard \fvarP$
is pure. This procedure should be done for each effect variable in $\Delta$
separately, so we generate fresh $\fvarP_\beta$ for each $\beta \in \Delta$.

\begin{remark}
  This trick works well for the problem effect inferece,
  because of flat, set-like structure of effects: any effect $\effect$
  containing variable $\alpha$ is equivalent to $\effect' \effjoin \alpha$
  for some $\effect'$ that doesn't contain $\alpha$.
  Similar property doesn't hold for types that have more rigid structure,
  so we don't see how our method could be
  applied for the problem of type inference.
\end{remark}

\paragraph {Subtyping.}

The \lstinline{subtype} function collects constraints that should be satisfied
in order for given subtyping judgement to hold.
But now, the constraints are represented in two forms:
set of subeffecting constraints~$\Omega$ and propositional-logic
formula~$\form$. Therefore, the \lstinline{subtype} function now returns
a pair of the form \lstinline[mathescape=true]{($\Omega$, $\form$)}.
As before, the only interesting case concerns
the effect-polymorphic quantifier.
\begin{lstlisting}[language=meta,mathescape=true]
subtype($\tforallE{\alpha} \type$; $\tforallE{\alpha} \type'$) =
  let ($\Omega$, $\form$) = subtype($\type$, $\type'$) in
  ($\Subst{\alpha}{\effpure}{\Omega}$; $
    \form \land \Omega \toFormula{\alpha}$)
\end{lstlisting}
Again after returning from the recursive call,
we need to prevent the variable~$\alpha$ from
escaping its scope while preserving the meaning of collected constraints.
First, we remove variable $\alpha$ from constraints set $\Omega$,
by substituting pure effect for $\alpha$.
The information lost by this step is restored in the form of
formula~$\Omega \toFormula{\alpha}$.

\paragraph{Constraint Separation.}

The other technical challenge in designing the algorithm comes from
restricting constraints that can appear in type schemes.
When examining let expression, the algorithm should divide constraints
from the first subexpressions into two sets:
one that can be included in the type scheme,
and that should be propagated upwards.
With the syntax of effects not extended with guards this task is easy.
Any constraint set can be normalized to the set of constraints
of the form $\subeffect{\alpha}{\effect}$~\cite{DBLP:conf/lomaps/NielsonNA96a},
and from such constraints we can select those that describe upper-bounds
for generalized effect variables.

In the case of extended syntax we can proceed similarly, if we slightly relax
the upper-bound condition.
First, observe that any constraint set can be normalized to set of
constraints of the form $\subeffect{\alpha\effguard\form}{\effect}$
(for details, we refer the reader to the appendix).
Then, for the purposes of the algorithm we allow such constraints
to appear in a type scheme if $\alpha$ is a variable bound by
the scheme.
For fixed valuation~$\rho$ of propositional variables, such a relaxation
doesn't influence on the expressive power of type schemes:
if $\rho\models\form$ then the constraint
$\subeffect{\alpha\effguard\form}{\effect}$
is equivalent to $\subeffect{\alpha}{\effect}$,
otherwise it is trivially satisfied.

Now, we can define function that separates constraints as follows.
\begin{lstlisting}[language=meta,mathescape=true]
separate($\Delta_g$; $\Omega$) =
  let $\Omega_n$ = normalize($\Omega$) in
  let $\Omega_g$ = $\{ \gamma \effguard \form \subtp \effect \mid
    (\gamma \effguard \form \subtp \effect) \in \Omega_n \land
    \gamma \in \Delta_g \}$ in
  let $\Omega_p$ = $\{
    \beta \effguard \form \subtp \bind{\subst{
      \overline{\gamma \in \Delta_g \mapsto \effpure}}} \effect \mid
    (\beta \effguard \form \subtp \effect) \in \Omega_n \land
    \beta \notin \Delta_g \}$ in
  ($\Omega_g$; $\Omega_p$)
\end{lstlisting}
The function takes set of variables $\Delta_g$ that will be generalized
in the scheme, and divieds set of constraints $\Omega$ into
the set $\Omega_g$ that will be included in the type scheme,
and the set $\Omega_p$ of remaining constraints.
The \lstinline{normalize} function normalizes constraints to the form
desribed above.
Additionally, we substitute pure effects for variables from $\Delta_g$
in $\Omega_p$, because $\Omega_p$ will be used outside the scope of $\Delta_g$.
By doing so, we don't lose any information:
all normalized constraints that non-trivially used variables from $\Delta_g$
are included in $\Omega_g$.

\paragraph{The algorithm.}

\begin{figure}[t!!]
\begin{lstlisting}[language=meta,mathescape=true]
infer($\Gamma$; $\lamE{\alpha} e$) =
  let ($P_1$; $\Delta_1$; $\type$; $\effect$; $\Omega_1$; $\form_1
    $) = infer($\Gamma$; $e$) in
  fresh $P_s$ = $\{ \fvarP_\beta \mid \beta \in \Delta_1 \}$ in
  fresh $\Delta_1'$ = $\{ \gamma_\beta \mid \beta \in \Delta_1 \}$ in
  let $\theta$ = $\subst{
    \overline{\beta \in \Delta_1 \mapsto \gamma_\beta
      \effjoin \alpha \effguard \fvarP_\beta}}$ in
  let $\Omega_1'$ = $\bind{\theta}
    (\Omega_1 \cup \{\effect \subtp \effpure\})$ in
  ($P_1 \uplus P_s$; $\Delta_1'$; $\tforallE{\alpha} \bind{\theta}\type
      $; $\effpure$; $
    \Subst{\alpha}{\effpure}{\Omega_1'}$; $
    \form_1 \land \Omega_1' \toFormula{\alpha}$)

infer($\Gamma$; $e\;[\seffect]$) =
  let ($P_1$; $\Delta_1$; $\type_1$; $\effect_1$; $\Omega_1$; $\form_1
    $) = infer($\Gamma$; $e$) in
  match $\type_1$ with
  | $\tforallE{\alpha} \type$ =>
    let ($\Delta_2$; $\effect_2$) = tr_effect($\seffect$) in
    ($P_1$; $\Delta_1 \uplus \Delta_2$; $
      \Subst{\alpha}{\effect_2}{\type}$; $\effect_1$; $
      \Omega_1$; $\form_1$)
  | _ => fail

infer($\Gamma$; {let} $x$ = $e_1$ {in} $e_2$) =
  let ($P_1$; $\Delta_1$; $\type_1$; $\effect_1$; $\Omega_1$; $\form_1
    $) = infer($\Gamma$; $e_1$) in
  fresh $\Delta_1'$ = $\{ \beta_\alpha \mid \alpha \in \Delta_1 \}$ in
  fresh $\Delta_g$ = $\{ \gamma_\alpha \mid \alpha \in \Delta_1 \}$ in
  let $\theta$ = $\subst{\overline{\alpha \in \Delta_1
    \mapsto \beta_\alpha \effjoin \gamma_\alpha}}$ in
  let ($\Omega_g, \Omega_p$) = separate($\Delta_g$; $
    \bind{\theta}(\Omega_1 \cup \{\effect_1 \subtp \effpure\})$) in
  let ($P_2$; $\Delta_2$; $\type_2$; $\effect_2$; $\Omega_2$; $\form_2
    $) = infer($
      \Gamma, x : \sforall{\Delta_g}{\Omega_g} \bind{\theta}\type_1$; $e_2$) in
  ($P_1 \uplus P_2$; $\Delta_1' \uplus \Delta_2$; $\type_2$; $\effect_2$; $
    \Omega_p \cup \Omega_2$; $\form_1 \land \form_2$)
\end{lstlisting}
  \caption{Selected cases of the effect inference algorithm.}
  \label{fig:algo-selected}
\end{figure}

Now we proceed to discussing the main function \lstinline{infer}
of the reconstruction algorithm.
As before, the function takes the environment $\Gamma$ and the examined
expression, but the returned tuple
\lstinline[mathescape=true]{($
  P$; $\Delta$; $\type$; $\effect$; $\Omega$; $\form$)}
contains two new components: a set~$P$ of generated propositional
variables, and a formula~$\form$ that can be seen as a kind of constraint.
Adding these components to the algorithm of \autoref{sec:simple} is
straightforward, so we focus on the cases presented in
\autoref{fig:algo-selected}, that require deeper explanation.
The full algorithm can be found in Appendix~\ref{appendix:algo}.

The effect abstraction $\lamE{\alpha}e$ introduces a
new effect variable~$\alpha$, so a special care should be taken
when the algorithm leaves the scope of $\alpha$.
First, the algorithm should decide for each variable
$\beta\in\Delta_1$ if it should be instantiated with effect containing $\alpha$
or not.
We do so the same way as we have seen in the \lstinline{tr_type} function:
by substituting $\gamma \effjoin \alpha \effguard \fvarP$ for $\beta$,
where $\fvarP$ is a fresh propositional variable representing the decision.
Moreover, we should transform constraints $\Omega_1'$ to the equivalent form
that doesn't contain variable~$\alpha$.
We proceed as we have seen in the \lstinline{subtype} function:
by substituting pure effect $\Subst{\alpha}{\effpure}{\Omega_1'}$,
and recovering lost information in the form of the formula
$\Omega_1' \toFormula{\alpha}$.

The effect application $e\appE{\seffect}$ doesn't pose any problems.
In this case, the algorithm makes sure that the type of the expression~$e$
is an effect-polymorphic quantifier~$\tforallE{\alpha}{\type}$,
and just substitutes for $\alpha$ the result of translating
syntactic effect~$\seffect$ into the internal representation.

For the let-expression,
the idea is simple: infer type $\type_1$ of the first expression,
generalize all generated variables in $\type_1$ together with constraints,
and continue with the second expression.
However two additional steps are required:
in the first step called \emph{variable splitting},
each variable $\alpha$ from $\Delta_1$ is split into
join of two variables $\beta_\alpha$ and $\gamma_\alpha$.
The former is propagated as one of generated variables,
while the latter is generalized in the polymorphic scheme.
In the second step, using the \lstinline{separate} function
constraints are separated into generalizable constraints~$\Omega_g$
in the relaxed upper-bound form, and remaining contraints~$\Omega_p$
that are propagated upwards.
The purpose of the second step was already explained,
when the \lstinline{separate} was described.

The purpose of variable splitting is to avoid too restrictive constraints
produced by the second step.
Intuitively, a variable $\alpha \in \Delta_1$ stands for
some unknown yet effect.
This effect can contain some variables
generalized in the scheme ($\gamma_\alpha$),
as well as other effects defined outside the let-expression.
The latter are unknown yet, so they are represented by
a variable $\beta_\alpha$.

\begin{example}
  To show that variable splitting is needed, consider a set
  $\{\texttt{IO} \subtp \alpha\}$ containing single constraint,
  where $\alpha$ was generated during the type inference.
  Without variable splitting, the \lstinline{separate} function would return
  a pair $\codetupII{\varnothing}{\{\texttt{IO} \subtp \effpure\}}$,
  where propagated contraints are obviously contradictory.
  On the other hand, with variable splitting we obtain a pair
  $\codetupII{\varnothing}{\{\texttt{IO} \subtp \beta_\alpha\}}$
  which preserves the original meaning of the input set.
\end{example}

\paragraph{Soundness and Completeness.}

The presented algorithm is sound and complete with respect to
the declarative system from the beginning of the section.

\begin{theorem}[Soundness]
  If $\codeinfer{\Gamma}{e} =
    \codetupVI{P}{\Delta'}{\type}{\effect}{\Omega}{\form}$
  for expression $e$ defined over $\Delta$,
  then
  $\RhoRelTyping{\Delta,\Delta'}{\Omega}{\Gamma}{e}{\type}{\effect}$
  for each valuation~$\rho$ satisfying formula~$\form$.
\end{theorem}

The Soundness Theorem is stated as precise as possible, therefore it uses
auxiliary typing relation for the extended syntax.
However, knowing the valuation~$\rho$ we can simplify all guarded effects
and return to the original syntax used by the declarative system.

\begin{corollary}
  If $\codeinfer{\Gamma}{e} =
    \codetupVI{P}{\Delta'}{\type}{\effect}{\Omega}{\form}$
  for expression $e$ defined over $\Delta$,
  then
  for each valuation~$\rho$ satisfying formula~$\form$
  there exist $\type'$, $\effect'$, and $\Omega'$ such that
  $\TypingRel{\Delta,\Delta'}{\Omega'}{\Gamma}{e}{\type'}{\effect'}$
  and $\RhoRel{\Delta,\Delta'}{\Omega}{\Omega'}$.
\end{corollary}

The statement of the completeness says that
the algorithm finds the most general solution,
\emph{i.e.,} that every derivation of the typing relation must be an instance
of the one found by the algorithm.

\begin{theorem}[Completeness]
  If $\TypingRel{\Delta}{\Omega}{\Gamma}{e}{\type}{\effect}$,
  then $\codeinfer{\Delta}{e} =
    \codetupVI{P'}{\Delta'}{\type'}{\effect'}{\Omega'}{\form}$
  for some $P'$, $\Delta'$, $\type'$, $\effect'$, $\Omega'$, and $\form$.
  Moreover, there exists substitution $\theta$
  that maps variables from $P'$ and $\Delta'$
  to formulae and effects, respectively, defined over $\Delta$, such that
    $\bind{\theta} \form$ is a tautology,
    $\EnvRel{\Delta}{\Omega}{\bind{\theta}\Omega'}$,
    \mbox{$\EnvRelSubType{\Delta}{\Omega}{\bind{\theta}\type'}{\type}$},
    and $\EnvRelSubEffect{\Delta}{\Omega}{\bind{\theta}\effect'}{\effect}$.
\end{theorem}

\subsection{Solving Toplevel Constraints}
\label{ssec:top-level}

The algorithm transforms the problem of effect inference to
a constraint-solving problem.
At the end, we get constraints of two kinds:
subeffecting constrainsts~$\Omega$ and a formula~$\form$.
We can solve these constraints in two steps.
\begin{enumerate}
\item
  Since we strictly avoid effect variables escaping their scopes,
  effect constraints in $\Omega$ can contain only top-level effect variables.
  We can pretend that all top-level effect variables are bound at the
  begining of the program, so we leave their scope as we proceeded
  in the effect-abstraction case.
  We transform the set $\Omega$ into formula
  $
    \form_\Omega \eqdef \bigwedge_{\alpha \in \Delta} \Omega \toFormula{\alpha}
  $,
  where $\Delta$ is a set of top-level effects.
  Note that the scope leaving procedure produces also constranits
  $\bind{\subst{\alpha \in \Omega \mapsto \effpure}} \Omega$,
  but since there are no variables in them, they are trivially satisfied.
\item
  The remaining work to do is
  to satisfy the formula $\form \wedge \form_\Omega$.
  This is a formula of propositional logic, so
  we can find a satisfying valuation using \emph{e.g.,} SAT~solver.
\end{enumerate}

\section{Type System Without Constraints}
\label{sec:no-constrs}

In this section we consider a slightly modified type system,
where type schemes are not allowed to contain constraints.
\begin{align*}
  \scheme &\Coloneqq \ssforall{\Delta}\type
\end{align*}
Since there is no way of abstracting constraints, the typing, subtyping,
and subeffecting judgements don't contain
the constraint environment~($\constrs$).
Here, we present only the two most notable inference rules of the
modified declarative system, as the others remain unchanged (except removing,
unnecessary constraint environment).
These rules resemble polymorphic instantiation and generalization
in the standard declarative formulation of the
Hindley-Milner type system~\cite{DBLP:journals/jcss/Milner78}.
\begin{mathpar}
  \inferrule
    { \Gamma(\varX)=\ssforall{\Delta'}\type}
    {\TypingRelS{\Delta}{\Gamma}
      {\varX}{\bind{\theta}\type}{\effpure}}

  \inferrule
    {\TypingRelS{\Delta,\Delta'}{\Gamma}{\expr_1}
        {\type_1}{\effpure} \and
      \TypingRelS{\Delta}
        {\Gamma, \varX:\ssforall{\Delta'}{\type_1}}{\expr_2}
        {\type_2}{\effect}}
    {\TypingRelS{\Delta}{\Gamma}
      {\elet{\varX}{\expr_1}{\expr_2}}{\type_2}{\effect}}
\end{mathpar}

Interestingly, the problem of effect inference in this system
is still decidable.  However, the price we pay is that
the algorithm generates a huge number of variables, which makes it
impractical for real-world programs.
Before we go into the details, we start with the two examples that illustrate
that the problem is not as simple as it may seem at first glance.

\begin{example}
  Since the let-binding
  (\lstinline[mathescape=true]{let $x$ = $e_1$ in $e_2$})
  introduces new effect variables, it might be tempting to
  abstract a single variable for each effect variable generated
  during the effect inference of $e_1$ and leave the scope similarly to
  the effect-abstraction case from the previous section.
  The let-binding case in the algorithm would then look like this:
  \begin{lstlisting}[language=meta,mathescape=true]
infer($\Gamma$; {let} $\varX$ = $e_1$ {in} $e_2$) =
  let ($P_1$; $\Delta_1$; $\type_1$; $\effect_1$; $\Omega_1$; $\form_1
    $) = infer($\Gamma$; $e_1$) in
  fresh $\Delta_1'$ = $\{ \beta_\alpha \mid \alpha \in \Delta_1 \}$ in
  fresh $\Delta_g$ = $\{ \gamma_\alpha \mid \alpha \in \Delta_1 \}$ in
  fresh $P_1'$ = $\{ \fvarP_\alpha \mid \alpha \in \Delta_1 \}$ in
  let $\theta$ = $\subst{\overline{\alpha \in \Delta_1
    \mapsto \beta_\alpha \effjoin \gamma_\alpha \effguard \fvarP_\alpha}}$ in
  let $\Omega_1'$ = $\bind{\theta}(\Omega_1 \cup \{\effect_1 \subtp \effpure\})$ in
  let ($P_2$; $\Delta_2$; $\type_2$; $\effect_2$; $\Omega_2$; $\form_2
    $) = infer($
      \Gamma, \varX : \ssforall{\Delta_g} \bind{\theta}\type_1
      $; $e_2$) in
  ($P_1 \uplus P_1' \uplus P_2$;$\Delta_1' \uplus \Delta_2$; $\type_2$; $\effect_2$; $
    \bind{\subst{\overline{\gamma_\alpha \mapsto \effpure}}}\Omega_1'
    \cup \Omega_2$; $
    \form_1 \land \form_2 \land
    \bigwedge_{\gamma \in \Delta_g}
      \Omega_1' \toFormula{\gamma}$)$\textrm{.}$
\end{lstlisting}
  This approach gives us a sound algorithm, but it is not complete.
  Assume that the algorithm on expression $e_1$ returns the
  tuple
  \begin{center}
  \lstinline[mathescape=true]{($P_1$; $\{\alpha_1, \alpha_2\}
    $; $\type_1(\alpha_1, \alpha_2)$; $\effpure
    $; $\{\alpha_1 \subtp \alpha_2, \alpha_2 \subtp \alpha_1\}$; $\form_1$)},
  \end{center}
  where $\type_1(\alpha_1, \alpha_2)$ is a type where $\alpha_1$ and $\alpha_2$
  occur in invariant positions.
  Since the generated constraints state that $\alpha_1$ and $\alpha_2$
  are equal, to obtain a completeness the algorithm should
  assign to variable $x$ a scheme equivalent to
  $\ssforall{\alpha}\type_1(\alpha, \alpha)$.
  However, the assigned scheme is
  \[
    \ssforall{\gamma_{\alpha_1}, \gamma_{\alpha_2}}
        \tau_1(\beta_{\alpha_1} \effjoin
                  \gamma_{\alpha_1} \effguard \fvarP_{\alpha_1},
               \beta_{\alpha_2} \effjoin
                  \gamma_{\alpha_2} \effguard \fvarP_{\alpha_2})\textrm{,}
  \]
  and the returned formula contains conjuncts
  $\theta(\alpha_1) \toFormula{\gamma_{\alpha_i}}
    \Rightarrow \theta(\alpha_2) \toFormula{\gamma_{\alpha_i}}$ for $i = 1, 2$,
  each of which simplifies to $(\beta_{\alpha_1}\effjoin\gamma_{\alpha_1}
    \effguard\fvarP_{\alpha_1})\toFormula{\gamma_{\alpha_i}}
    \Rightarrow (\beta_{\alpha_2}\effjoin\gamma_{\alpha_2}\effguard
    \fvarP_{\alpha_2})\toFormula{\gamma_{\alpha_i}}$, and further to
  $\fvarP_{\alpha_1}\Rightarrow\bot$ and $\fvarP_{\alpha_2}\Rightarrow\bot$.
  The obtained scheme is not polymorphic at all, so such a naive algorithm
  is not complete.
\end{example}

From the above example we see that the algorithm should take into account
the fact that generated effect variables are not necessarily independent.
We can achieve this by allowing any combination of generalized
variables ($\gamma$) to be substituted for the generated effect variables
($\alpha$). The change is relatively simple:
we substitute for $\alpha$ a join of all the effect variables from $\Delta_g$,
guarded by fresh propositional variables.
The rest of the algorithm remains unchanged.
\begin{lstlisting}[language=meta,mathescape=true]
infer($\Gamma$; {let} $x$ = $e_1$ {in} $e_2$) =
  ...
  fresh $\Delta_g$ = ... in
  fresh $P_1'$ = $\{ \fvarP_{\alpha,\gamma} \mid \alpha \in \Delta_1
    \land \gamma \in \Delta_g \}$ in
  let $\theta$ = $\overline{\alpha \in \Delta_1
    \mapsto \beta_\alpha \effjoin
      \bigeffjoin{\gamma\in\Delta_g} \gamma \effguard \fvarP_{\alpha,\gamma}}$ in
  ...
\end{lstlisting}

Since we take any combination of variables from $\Delta_g$,
the size of $\Delta_g$ doesn't have to be the same as the size of $\Delta_1$.
The question is how many variables we need to generate in $\Delta_g$ to
obtain a complete algorithm. The next example shows that sometimes we need
to generate strictly more variables than the size of $\Delta_1$.

\begin{example}
  Assume that the size of $\Delta_g$ is equal to the size of $\Delta_1$.
  The above algorithm is sound, but still not complete.
  To present a counterexample,
  assume that the algorithm called on subexpression~$e_1$ of
  let binding \lstinline[mathescape=true]+let $x$ = $e_1$ in $e_2$+~returns
  \begin{center}
  \lstinline[mathescape=true]{($P_1$; $\{\alpha_0, \alpha_1, \alpha_2\}
    $; $\type_1(\alpha_0, \alpha_1, \alpha_2)$; $\effpure
    $; $\{\alpha_0 \subtp \alpha_1 \effjoin \alpha_2\}$; $\form_1$)},
  \end{center}
  where variables $\alpha_0$, $\alpha_1$, and $\alpha_2$ occur
  on invariant position in $\tau_1(\alpha_0, \alpha_1, \alpha_2)$.
  Moreover, suppose that $e_2$ is typeable only if
  both
  \begin{equation*}
    \type_p \eqdef \type_1(\effpure, \effpure, \effpure) \qquad
    \type_i \eqdef \type_1(A_1 \effjoin A_2, A_1 \effjoin B_1, A_2 \effjoin B_2)
  \end{equation*}
  are valid instantiations of a type scheme assigned to $x$
  (where $A_1$, $A_2$, $B_1$, $B_2$ are some pairwise different effect
  constants).
  The type-scheme assigned to~$x$ by the algorithm would be
  \begin{align*}
    \sigma \eqdef \ssforall{\gamma_0, \gamma_1, \gamma_2} \type_1(
      & \beta_{\alpha_0} \effjoin
        \gamma_0 \effguard \fvarP_{0,0} \effjoin
        \gamma_1 \effguard \fvarP_{0,1} \effjoin
        \gamma_2 \effguard \fvarP_{0,2},\\
      & \beta_{\alpha_1} \effjoin
        \gamma_0 \effguard \fvarP_{1,0} \effjoin
        \gamma_1 \effguard \fvarP_{1,1} \effjoin
        \gamma_2 \effguard \fvarP_{1,2},\\
      & \beta_{\alpha_2} \effjoin
        \gamma_0 \effguard \fvarP_{2,0} \effjoin
        \gamma_1 \effguard \fvarP_{2,1} \effjoin
        \gamma_2 \effguard \fvarP_{2,2})
  \end{align*}
  and the formula generated from the constraint
  $\alpha_0 \subtp \alpha_1\effjoin\alpha_2$
  contains a conjunction of the following subformulae.
  \[
    \fvarP_{0,0}  \Rightarrow \fvarP_{1,0} \vee \fvarP_{2,0} \qquad
    \fvarP_{0,1}  \Rightarrow \fvarP_{1,1} \vee \fvarP_{2,1} \qquad
    \fvarP_{0,2}  \Rightarrow \fvarP_{1,2} \vee \fvarP_{2,2}
  \]
  An inquisitive reader may verify that there is no valuation of propositional
  variables that satisfies the above formulae and allows
  $\type_p$ and $\type_i$ to be valid instantiations of~$\sigma$.
  On the other hand, the expression is typeable in the declarative type system
  if $x$ gets the type scheme
  \begin{align*}
    \sigma' & \eqdef \ssforall{\gamma_0, \gamma_1, \gamma_2, \gamma_3}
      \type_1(\gamma_0 \effjoin \gamma_1,
             \gamma_0 \effjoin \gamma_2,
             \gamma_1 \effjoin \gamma_3)\textrm{.}
  \end{align*}
\end{example}

The above example shows that when the size of $\Delta_g$ is the same as $\Delta_1$
the algorithm is still not complete.
However, it gives us hope that if we are able to find an upper-bound
for the number of generalized variables used by the declarative system (in this
example four is enough), completeness could be regained.
Observe that in a type scheme $\ssforall{\alpha_1,\ldots,\alpha_n}\type$ if
$n > 2^{\sizeof{\type}}$,
where $\sizeof{\type}$ is the number of arrows in $\type$,
then there are two variables $\alpha_i$ and $\alpha_j$
that always appear together in $\type$.
We could equate them and get the equivalent scheme
$
\ssforall{\alpha_1,\ldots\alpha_{i-1},\alpha_{i+1},\alpha_n}
  \bind{\subst{\alpha_i\mapsto\alpha_j}}\type
$.
\begin{example}
  A type scheme
  $\ssforall{\beta,\gamma}\alpha \tarrow{\beta\effjoin\gamma} \alpha$
  is equivalent to $\ssforall{\beta}\alpha \tarrow{\beta} \alpha$, \emph{i.e.,}
  they have the same instantiations.
\end{example}

Moreover, $\sizeof{\type}$ does not depend on effects (and is invariant
with respect to effect substitutions), therefore it is known after the
type reconstruction phase! 
With this intuitions in mind, we can construct the algorithm that is sound
and complete with respect to the declarative system and uses
$2^{\sizeof{\type}}$ as a mentioned upper-bound.
In sake of brevity, we present only the let-case.
\begin{lstlisting}[language=meta,mathescape=true]
infer($\Gamma$; {let} $x$ = $e_1$ {in} $e_2$) =
  let ($P_1$; $\Delta_1$; $\type_1$; $\effect_1$; $\Omega_1$; $\form_1
    $) = infer($\Gamma$; $e_1$) in
  fresh $\Delta_1'$ = $\{ \beta_\alpha \mid \alpha \in \Delta_1 \}$ in
  fresh $\Delta_g$ = $\{ \gamma_i \mid i \in 1,\ldots,2^{\sizeof{\tau_1}} \}$ in
  fresh $P_1'$ = $\{ \fvarP_{\alpha,\gamma} \mid \alpha \in \Delta_1
    \land \gamma \in \Delta_g \}$ in
  let $\theta$ = $\subst{\overline{\alpha \in \Delta_1
    \mapsto \beta_\alpha \effjoin
      \bigeffjoin{\gamma\in\Delta_g} \gamma \effguard \fvarP_{\alpha,\gamma}}}$ in
  let $\Omega_1'$ = $\bind{\theta}(\Omega_1 \cup \{\effect_1 \subtp \effpure\})$ in
  let ($P_2$; $\Delta_2$; $\tau_2$; $\effect_2$; $\Omega_2$; $\form_2
    $) = infer($
      \Gamma, x : \ssforall{\Delta_g} \bind{\theta}\type_1
      $; $e_2$) in
  ($P_1 \uplus P_1' \uplus P_2$;$\Delta_1' \uplus \Delta_2$; $\type_2$; $\effect_2$; $
    \bind{\subst{\overline{\gamma_\alpha \mapsto \effpure}}}\Omega_1'
    \cup \Omega_2$; $
    \form_1 \land \form_2 \land
    \bigwedge_{\gamma \in \Delta_g}
      \Omega_1' \toFormula{\gamma}$)$\textrm{.}$
\end{lstlisting}

\begin{theorem}[Soundness]
  Suppose that $\codeinfer{\Gamma}{e} =
    \codetupVI{P}{\Delta'}{\type}{\effect}{\Omega}{\form}$
  for some expression $e$ defined over $\Delta$ and
  $\theta$ is a substitution mapping $P$ and $\Delta, \Delta'$ to
  terms over $\Delta''$.
  Then $\Delta'' \vdash_{\!\rho} \bind{\theta}\Omega$ implies
  $\RhoRelTyping{\Delta''}{\bind{\theta}\Omega}{\bind{\theta}\Gamma}{e}{\bind{\theta}\type}{\bind{\theta}\effect}$
  for each valuation~$\rho$ satisfying formula~$\bind{\theta}\form$.
\end{theorem}

\begin{theorem}[Completeness]
  If $\TypingRelS{\Delta}{\Gamma}{e}{\type}{\effect}$,
  then $\codeinfer{\Delta}{e} =
    \codetupVI{P'}{\Delta'}{\type'}{\effect'}{\Omega'}{\form}$
  for some $P'$, $\Delta'$, $\type'$, $\effect'$, $\Omega'$, and $\form$.
  Moreover, there exists substitution $\theta$
  that maps variables from $P'$ and $\Delta'$
  to formulae and effects, respectively, defined over $\Delta$, such that
    $\bind{\theta} \form$ is a tautology,
    $\Delta \vdash \bind{\theta}\Omega'$,
    $\EnvRelSubTypeS{\Delta}{\bind{\theta}\type'}{\type}$,
    and $\EnvRelSubEffectS{\Delta}{\bind{\theta}\effect'}{\effect}$.
\end{theorem}

\section{Formalization}
\label{sec:formalization}

The metatheory of effect inference in presence of general polymorphism
is particularly prone to errors related to variable bindings.
In order to avoid such mistakes, we formalized all the presented results
using Rocq proof assistant.
The key issue in such a formalization is the representation of
variable binding.
We decided to use
functorial approach~\cite{DBLP:conf/lics/FiorePT99} a.k.a. nested datatypes
approach~\cite{DBLP:conf/mpc/BirdM98}.
However, the algorithm strongly relies on the finiteness of the generated
sets,
so parametrizing syntax with arbitrary types instead of finite sets,
like in the original nested datatype approach does not work well.

At the beginning we tried to use Binding library~\cite{ws/coqpl/PolesiukS24}
designed for a functorial approach, because of its flexibility in
the representation of sets of potentially free variables.
We parametrized the syntax with a cartesian product of
finite sets with a disjoint union as one of constructors.
However, Binding heavily uses type classes mechanism that didn't
work well when multiple instantiations of the same class were used
in a single formalization.
Therefore, we ended up with a functorial approach reimplemented from scratch.

\section{Implementation and Practical Considerations}
\label{sec:implementation}

Looking at the algorithm presented in this work, it might not be immediately
obvious whether it is viable in practice. In the following, we will share
some insights we have gained from implementing it in our \langname{}
programming language.

\paragraph{Higher kinds.}
While our implementation includes arrow kinds $\kind_1 \to \kind_2$, it has
the restriction that the effect kind cannot appear on the right-hand
side of an arrow. Due to this, type-level application cannot appear in
effects, and so effects remain simple sets of variables. As we have not
found any use cases for effects depending on types or effects in our
language, we consider this limitation to be~acceptable.

\paragraph{Constraint simplification.}
The constraints generated by the algorithm can be quite large if left as-is.
In turn, the schemes inferred for let definitions run the risk of containing
huge constraint sets, which are unwieldy for the programmer to read and
inefficient for the implementation to satisfy at each use site of such a
definition.
Fortunately, in practice, constraints can often be significantly simplified,
or completely eliminated, with a collection of straightforward heuristics,
for example similar to those in \cite{DBLP:conf/lomaps/NielsonNA96a}.

\paragraph{Checking formula satisfiability.}
After running our algorithm and eliminating the resulting subeffect
constraints, we are left with a propositional-logic formula which needs
to be satisfied in order to consider the effect inference successful.
Though of course the SAT problem is well-known to be NP-complete,
in practice we found that the formulae generated by the algorithm
are easy to solve. In the current implementation, we use a hand-written
SAT solver that exploits certain properties of the formulae to attain a
reasonable runtime, but since many optimized SAT solvers are available,
a ready-made solution is also an option.

\paragraph{REPL and incremental solving.}
When running the interpreter in REPL mode, it is necessary to check the
satisfiability of the produced formula for each expression
and definition given by the programmer to catch any errors. However,
unlike when operating on complete programs, the values of propositional
variables cannot be fixed too eagerly, as future input can make some
valuations invalid, while other satisfying valuations remain.
The simplest implementation can check whether any satisfying valuation
exists after each input, but not set the values of any propositional variables.
Unfortunately, that results in checking an ever-larger formula every time,
which could be a problem for long-running REPL sessions.
As an optimization, in our implementation we fix the values of propositional
variables as soon as possible if there is only one satisfying valuation for
them.

\section{Related Work}

\paragraph{Effect inference in presence of higher-rank polymorphism.}

Jouvelot and Gifford~\cite{DBLP:conf/popl/JouvelotG91}
presented an algorithm of effect inference in their
FX programming language~\cite{%
  DBLP:conf/popl/LucassenG88,%
  Gifford1992ReportOT}.
Their algorithm collects effect equivalence constraints
and attaches them to algebraic type schemes.
Their language supports explicit higher-rank
polymorphism and allows omitting type annotation in $\lambda$-abstractions,
but doesn't support subtyping and wildcards under effect quantifiers.
This is the only work that we are aware of that tackles the problem of
effect reconstruction in a setup similar to ours.
Jouvelot and Gifford claimed that their algorithm is sound and complete.
However, it turned out that they had a subtle bug related to variable binding
that made their theorems untrue
with no simple fix in sight~\cite{Jouvelot-communication}.

\paragraph{Effect inference as a static analysis.}

After the original work on effect inference by Jouvelot and Gifford, the
research community's attention shifted in the direction of static analysis.
Talpin and Jouvelot~\cite{DBLP:journals/jfp/TalpinJ92}
used the technique of algebraic type schemes
to perform effect and region inference.
Because the region inference is more like a static analysis transparent
to the user, considering explicit higher-rank polymorphism doesn't
make much sense, so they focused on ML-style polymorphism only,
making their work free of mentioned bug.
This work started a long line of research
on effect based static analysis~\cite{%
  DBLP:journals/iandc/TalpinJ94,%
  DBLP:conf/popl/TofteT94,%
  DBLP:conf/ccl/NielsonN94,%
  tang1994control,%
  DBLP:conf/popl/BirkedalTV96,%
  DBLP:conf/lomaps/NielsonNA96a,%
  DBLP:conf/sas/FahndrichA97,%
  DBLP:journals/toplas/TofteB98,%
  DBLP:conf/fase/NielsonAN98,%
  DBLP:journals/toplas/HelsenT04,%
  DBLP:journals/pacmpl/Elsman24}.
Nielson and Nielson~\cite{nielson2000type}
observed that subtyping doesn't influece shape of types,
and concluded that this allows for two-stage approach.
Amtoft et~al.~\cite{DBLP:journals/jfp/AmtoftNN97}
and Birkedal and Tofte~\cite{DBLP:journals/tcs/BirkedalT01}
attacked the problem of effect inference where type schemes doesn't contain
constraints. However, their systems doesn't admit general polymorhism,
and we don't see how their solutions may scale to our system.

\paragraph{Effect inference in surface type systems.}

The interest in type-and-effect systems as a facility exposed to the
programmer has experienced a resurgence alongside the research on algebraic
effects~\cite{%
  DBLP:journals/entcs/PlotkinP04,%
  DBLP:journals/corr/PlotkinP13,%
  DBLP:journals/corr/Leijen14,%
  DBLP:conf/haskell/WuSH14,%
  DBLP:journals/jlp/BauerP15,%
  DBLP:conf/icfp/HillerstromL16,%
  DBLP:journals/pacmpl/BiernackiPPS18,%
  DBLP:journals/pacmpl/ZhangM19,%
  DBLP:journals/pacmpl/BrachthauserSO20,%
  DBLP:journals/pacmpl/BiernackiPPS20,%
  DBLP:journals/pacmpl/XieCIL22,%
  DBLP:conf/onward/Madsen22,%
  DBLP:conf/esop/VilhenaP23}.
However, the idea is much older, and the previously discussed work by Jouvelot
and Gifford~\cite{DBLP:conf/popl/JouvelotG91} fits into this category.
Surprisingly, later research in this area does not build on that work, and most
of it opts for inference based on row
reconstruction~\cite{%
  DBLP:conf/lics/Wand87,%
  10.5555/186677.186689}.

The idea to use row polymorphism for effects first appeared in
Links~\cite{DBLP:conf/tldi/LindleyC12}. The advantage of this approach
is that since row unification is decidable, it is easy to extend the
standard Hindley-Milner algorithm with effect rows while maintaining
completeness of the inference. Effect rows are essentially lists
of behaviors, possibly ending with a row variable.
There is a lot of work on effect rows~\cite{%
  DBLP:conf/tldi/LindleyC12,%
  DBLP:conf/icfp/HillerstromL16,%
  DBLP:journals/corr/Leijen14,%
  DBLP:conf/sfp/IkemoriCML22,%
  DBLP:conf/esop/VilhenaP23}
with slightly different approaches and design decisions.
From our perspective, the main disadvantage of effect rows is that they
are conceptually more complex than sets, and also less
intuitive. Additionally, some expressiveness is lost by not being able to
use multiple polymorphic variables within a row.

Two notable exceptions with effect sets rather than rows include the
Flix~\cite{DBLP:conf/onward/Madsen22} and
Effekt~\cite{DBLP:journals/pacmpl/BrachthauserSO20} languages.
In Flix, effects are sets with all the usual boolean operations such as
union, intersection and difference. As a result, effect inference can be
implemented using boolean
unification~\cite{DBLP:journals/pacmpl/MadsenP20}.
We note the common thread between our solution and the approach used by Flix:
both reduce inference to solving boolean algebra problems of some kind, in our
case satisfying propositional formulae, and in the case of Flix performing
boolean unification on sets.
We found the rich grammar of effects in Flix a bit too complicated for our
needs, as we sought to present the programmer with effects with just
the union operation. Effekt takes an entirely different
approach by replacing parametric effect
polymorphism with contextual polymorphism. In summary, this view of
polymorphism means that effects do not need to contain polymorphic variables at
all, and in turn effects can be represented as sets without complicating
effect inference. This simplicity comes with a trade-off:
functions in Effekt are second-class.

\paragraph{Type inference in System F.}

Full type reconstruction of System~F is known to be undecidable~\cite{%
DBLP:journals/apal/Wells99}, however algorithms to reconstruct types in presence
of partial annotation have been developed for some time. Starting with Pierce
\emph{et al.}~\cite{DBLP:journals/toplas/PierceT00}, bidirectional approach to
this problem became a standard. Here community split into two paths.
One dedicated to problem of type checking in System~F with explicit type
application~\cite{%
  DBLP:conf/icfp/DunfieldK13,%
  DBLP:conf/pldi/EmrichLSCC20,%
  DBLP:journals/pacmpl/DunfieldK19,%
  DBLP:conf/icfp/BotlanR03}
The other decided to loosen the type system, and adopted rank-N
polymorphism~\cite{DBLP:conf/popl/OderskyL96,DBLP:journals/jfp/JonesVWS07}.

\paragraph{Formalization of type reconstruction.}

Surprisingly, type reconstruction algorithms were rarely formalized
using proof assistants for a long time.
In the early works~\cite{%
    DBLP:journals/jar/NaraschewskiN99,%
    DBLP:journals/jar/DuboisM99,%
    Urban_Nipkow_2009,%
    DBLP:conf/aplas/Garrigue10}
a concrete or nominal approach to variable binding was used,
and the scopes of variables were managed manually with relatively
large overhead.
The situation has changed when Dunfield and Krishnaswami~\cite{%
  DBLP:conf/icfp/DunfieldK13} proposed a framework for type reconstruction
algorithms where scopes of unification variables are tracked via ordered
contexts. They didn't formalized their work in a proof assistant,
but following their approach many advanced algorithms related to subtyping
were formalized in Abela~\cite{%
  DBLP:conf/itp/ZhaoOS18,%
  DBLP:journals/pacmpl/ZhaoOS19,%
  DBLP:conf/ecoop/ZhaoO22,%
  DBLP:journals/pacmpl/CuiJO23%
  }, Coq/Rocq~\cite{%
  DBLP:conf/itp/BosmanKS23,%
  DBLP:journals/pacmpl/JiangCO25%
  }, and Agda~\cite{DBLP:journals/pacmpl/XueO24}.
Our result, also explicitly keeps track of variable scopes, but instead of
using ordered contexts, we rely on functorial syntax.
Recently, Fan \emph{et~al.}~\cite{DBLP:journals/pacmpl/FanXX25} presented
a different approach to formalizing scopes using levels, which is closer
to modern implementations of type reconstruction algorithms.

\section{Conclusion and Future Work}

In this paper we have proposed an effect reconstruction algorithm with
set-like semantics of effects and support for higher-rank polymorphism.
The algorithm requires only minimal annotations from the programmer.
Our algorithm utilizes effect guards and extracts propositional formulae
from subeffecting constraints in order to correctly manage the scopes of
effect variables.
We have proven our algorithm to be both sound and complete with respect
to the declarative type-and-effect system.
We have also shown that it is feasible to implement in practice by adding it to
a realistic programming language with algebraic effects.

This study opens a number of avenues for further research.
One concerns the algorithm without constraints in schemes presented in
\autoref{sec:no-constrs}.
Our upper bound for the number of variables needed to preserve
the algorithm's completeness is $2^{\sizeof{\tau}}$.
However this estimation does not take into account the number of constraints
associated with the variables present in the type.
It is easy to see that in the case where there are no such constraints, there
is no need to create any fresh variables. This observation suggests that a
better upper bound exists, which gives the hope for a practical implementation.

In the opposite direction, we can imagine allowing the programmer
to provide algebraic type schemes in rank-N annotations.
For example, this would enable types like
$\tforall{\tvarA}
  (\sforall{\tvarB}{\subeffect{\tvarB}{\tvarA}} \texttt{Int}\to_{\seffwildcard} \texttt{Int})
  \to_{\seffwildcard} \texttt{Int}$.
We believe it will be possible to have sound and complete algorithm in such
a setting, but foresee difficulties associated with calculating the
transitive closure of subeffecting constraints.

\bibliographystyle{splncs04}
\bibliography{references}

\clearpage
\appendix
\section{Declarative System}
\label{appendix:system}

\reldefline{Effect Matching}
  {$\EnvRelMatch{\Delta}{\seffect}{\effect}$}
\begin{mathpar}
  \inferrule{ }{\EnvRelMatch{\Delta}{\tvarKE}{\tvarKE}}

  \inferrule{ }{\EnvRelMatch{\Delta}{\effpure}{\effpure}}

  \inferrule{
    \EnvRelMatch{\Delta}{\seffect_1}{\effect_1}
    \and
    \EnvRelMatch{\Delta}{\seffect_2}{\effect_2}}
    {\EnvRelMatch{\Delta}{\seffect_1 \effjoin \seffect_2}
      {\effect_1 \effjoin \effect_2}}

  \inferrule{ }{\EnvRelMatch{\Delta}{\seffwildcard}{\effect}}
\end{mathpar}

\reldefline{Type Matching}
  {$\EnvRelMatch{\Delta}{\stype}{\type}$}
\begin{mathpar}
  \inferrule{ }{\EnvRelMatch{\Delta}{\tvarKT}{\tvarKT}}

  \inferrule{
    \EnvRelMatch{\Delta}{\stype_1}{\type_1}
    \and
    \EnvRelMatch{\Delta}{\stype_2}{\type_2}
    \and
    \EnvRelMatch{\Delta}{\seffect}{\effect}}
    {\EnvRelMatch{\Delta}{\stype_1 \tarrow{\seffect} \stype_2\ }
      {\ \type_1 \tarrow{\effect} \type_2}}

  \inferrule{
    \EnvRelMatch{\Delta, \alpha^\ktype}{\stype}{\type}}
    {\EnvRelMatch{\Delta}{\tforallT{\alpha} \stype}{\tforallT{\alpha} \type}}

  \inferrule{
    \EnvRelMatch{\Delta, \alpha^\keffect}{\stype}{\type}}
    {\EnvRelMatch{\Delta}{\tforallE{\alpha} \stype}{\tforallE{\alpha} \type}}
\end{mathpar}

\reldefline{Subeffecting}
  {$\EnvRelSubEffect{\Delta}{\constrs}{\effect_1}{\effect_2}$}
\begin{mathpar}
  \inferrule{ }{\EnvRelSubEffect{\Delta}{\constrs}{\effect}{\effect}}

  \inferrule{
    \EnvRelSubEffect{\Delta}{\constrs}{\effect_1}{\effect_2}
    \and
    \EnvRelSubEffect{\Delta}{\constrs}{\effect_2}{\effect_3}}
    {\EnvRelSubEffect{\Delta}{\constrs}{\effect_1}{\effect_3}}

  \inferrule{
    (\subeffect{\effect_1}{\effect_2}) \in \constrs}
    {\EnvRelSubEffect{\Delta}{\constrs}{\effect_1}{\effect_2}}

  \inferrule{ }{\EnvRelSubEffect{\Delta}{\constrs}{\effpure}{\effect}}

  \inferrule{
    \EnvRelSubEffect{\Delta}{\constrs}{\effect_1}{\effect}
    \and
    \EnvRelSubEffect{\Delta}{\constrs}{\effect_2}{\effect}}
    {\EnvRelSubEffect{\Delta}{\constrs}{\effect_1 \effjoin \effect_2}{\effect}}
  \\
  \inferrule{
    \EnvRelSubEffect{\Delta}{\constrs}{\effect}{\effect_1}}
    {\EnvRelSubEffect{\Delta}{\constrs}{\effect}{\effect_1 \effjoin \effect_2}}

  \inferrule{
    \EnvRelSubEffect{\Delta}{\constrs}{\effect}{\effect_2}}
    {\EnvRelSubEffect{\Delta}{\constrs}{\effect}{\effect_1 \effjoin \effect_2}}
\end{mathpar}

\reldefline{Subtyping}
  {$\EnvRelSubType{\Delta}{\constrs}{\type_1}{\type_2}$}
\begin{mathpar}
  \inferrule{ }{\EnvRelSubType{\Delta}{\constrs}{\tvarKT}{\tvarKT}}

  \inferrule{
    \EnvRelSubType{\Delta}{\constrs}{\type_2'}{\type_1'}
    \and
    \EnvRelSubType{\Delta}{\constrs}{\type_1}{\type_2}
    \and
    \EnvRelSubEffect{\Delta}{\constrs}{\effect_1}{\effect_2}}
    {\EnvRelSubType{\Delta}{\constrs}
      {\type_1' \tarrow{\effect_1} \type_1}
      {\type_2' \tarrow{\effect_2} \type_2}}

  \inferrule{
    \EnvRelSubType{\Delta, \alpha^\ktype}{\constrs}{\type_1}{\type_2}}
    {\EnvRelSubType{\Delta}{\constrs}
      {\tforallT{\alpha}{\type_1}}
      {\tforallT{\alpha}{\type_2}}}

  \inferrule{
    \EnvRelSubType{\Delta, \alpha^\keffect}{\constrs}{\type_1}{\type_2}}
    {\EnvRelSubType{\Delta}{\constrs}
      {\tforallE{\alpha}{\type_1}}
      {\tforallE{\alpha}{\type_2}}}
\end{mathpar}

\reldefline{Typing}
  {$\TypingRel{\Delta}{\Gamma}{\constrs}{\expr}{\type}{\effect}$}
\begin{mathpar}
  \inferrule{
    \Gamma(\varX)=\sforall{\Delta'}{\constrs'}\type
    \and
    \Delta;\constrs\vdash\bind{\theta}\constrs'}
    {\TypingRel{\Delta}{\constrs}{\Gamma}
      {\varX}{\bind{\theta}\type}{\effpure}}
  \hspace{2em}
  \inferrule{
    \EnvRelMatch{\Delta}{\stype}{\type_1}
    \and
    \TypingRel{\Delta}{\constrs}{\Gamma,\varX:\type_1}
        {\expr}{\type_2}{\effect}}
    {\TypingRel{\Delta}{\constrs}{\Gamma}{\lam{\varX}{\stype}\expr}
      {\type_1\to_{\effect}\type_2}{\effpure}}

  \inferrule{
    \TypingRel{\Delta,\tvarA^\ktype}{\constrs}
      {\Gamma}{\expr}{\type}{\effpure}}
    {\TypingRel{\Delta}{\constrs}{\Gamma}{\lamT{\tvarA}\expr}
      {\tforallT{\tvarA}\type}{\effpure}}

  \inferrule{
    \TypingRel{\Delta,\tvarA^\keffect}{\constrs}
      {\Gamma}{\expr}{\type}{\effpure}}
    {\TypingRel{\Delta}{\constrs}{\Gamma}{\lamE{\tvarA}\expr}
      {\tforallE{\tvarA}\type}{\effpure}}

  \inferrule{
    \TypingRel{\Delta}{\constrs}{\Gamma}{\expr_1}
        {\type_2\to_{\effect}\type_1}{\effect}
    \and
    \TypingRel{\Delta}{\constrs}{\Gamma}{\expr_2}
        {\type_2}{\effect}}
    {\TypingRel{\Delta}{\constrs}{\Gamma}{\expr_1\;\expr_2}
      {\type_1}{\effect}}

  \inferrule{
    \TypingRel{\Delta}{\constrs}{\Gamma}{\expr}
        {\tforallT{\tvarA}{\type}}{\effect}
    \and
    \EnvRelMatch{\Delta}{\stype}{\type'}}
    {\TypingRel{\Delta}{\constrs}{\Gamma}{\expr \appT{\stype}}
      {\Subst{\tvarA}{\type'}{\type}}{\effect}}

  \inferrule{
    \TypingRel{\Delta}{\constrs}{\Gamma}{\expr}
        {\tforallE{\tvarA}{\type}}{\effect}
    \and
      \EnvRelMatch{\Delta}{\seffect}{\effect'}}
    {\TypingRel{\Delta}{\constrs}{\Gamma}{\expr \appE{\seffect}}
      {\Subst{\tvarA}{\effect'}{\type}}{\effect}}

  \inferrule{
    \TypingRel{\Delta,\Delta^\keffect_g}{\constrs,\constrs_g}
        {\Gamma}{\expr_1}{\type_1}{\effpure}
    \and
    \TypingRel{\Delta}{\constrs}
      {\Gamma, \varX:\sforall{\Delta^\keffect_g}{\constrs_g}{\type_1}}{\expr_2}
        {\type_2}{\effect}}
    {\TypingRel{\Delta}{\constrs}{\Gamma}
      {\elet{\varX}{\expr_1}{\expr_2}}{\type_2}{\effect}}

  \inferrule{
    \TypingRel{\Delta}{\constrs}{\Gamma}{\expr}{\type}{\effect} \\
      \EnvRelSubType{\Delta}{\constrs}{\type}{\type'} \\
      \EnvRelSubEffect{\Delta}{\constrs}{\effect}{\effect'}}
    {\TypingRel{\Delta}{\constrs}{\Gamma}{\expr}{\type'}{\effect'}}
\end{mathpar}

\section{The Algorithm}
\label{appendix:algo}

\begin{lstlisting}[language=meta,mathescape=true]
tr_effect($\tvarA$) = ($\varnothing$; $\tvarA$)
tr_effect($\effpure$) = ($\varnothing$; $\effpure$)
tr_effect($\seffect_1 \effjoin \seffect_2$) =
  let ($\Delta_1$; $\effect_1$) = tr_effect($\seffect_1$) in
  let ($\Delta_2$; $\effect_2$) = tr_effect($\seffect_2$) in
  ($\Delta_1 \uplus \Delta_2$; $\effect_1 \effjoin \effect_2$)
tr_effect($\seffwildcard$) =
  fresh $\alpha$ in
  ($\{\alpha\}$; $\alpha$)
\end{lstlisting}

\begin{lstlisting}[language=meta,mathescape=true]
tr_type($\alpha$) = ($\varnothing$; $\varnothing$; $\alpha$)
tr_type($\tforallT{\alpha} \stype$) =
  let ($P$; $\Delta$; $\type$) = tr_type($\stype$) in
  ($P$; $\Delta$; $\tforallT{\alpha} \type$)
\end{lstlisting}

\begin{lstlisting}[language=meta,mathescape=true]
tr_type($\tforallE{\alpha} \stype$) =
  (* $\alpha$ is fresh, because we open the binder *)
  let ($P$; $\Delta$; $\type$) = tr_type($\stype$) in
  fresh $P_s$ = $\{ \fvarP_\beta \mid \beta \in \Delta \}$ in
  fresh $\Delta'$ = $\{ \gamma_\beta \mid \beta \in \Delta \}$ in
  ($P \uplus P_s$; $\Delta'$; $\tforallE{\alpha} \bind{\subst{
    \overline{\beta \in \Delta \mapsto \gamma_\beta
      \effjoin \alpha \effguard \fvarP_\beta}}} \type $)
\end{lstlisting}

\begin{lstlisting}[language=meta,mathescape=true]
tr_type($\stype_1 \tarrow{\seffect} \stype_2$) =
  let ($P_1$; $\Delta_1$; $\type_1$) = tr_type($\stype_1$) in
  let ($P_2$; $\Delta_2$; $\type_2$) = tr_type($\stype_2$) in
  let ($\Delta_3$; $\effect$) = tr_effect($\seffect$) in
  ($P_1 \uplus P_2$; $\Delta_1 \uplus \Delta_2 \uplus \Delta_3$; $
    \type_1 \tarrow{\effect} \type_2$)
\end{lstlisting}

As a specification of \lstinline{tr_type} function we can prove the
following two lemmas, that state its soundness and completeness.

\begin{lemma}
  If $\codeTrType{\stype} = \codetupIII{P}{\Delta'}{\type}$
  for $\stype$ defined over $\Delta$,
  then $\RhoRelMatch{\Delta,\Delta'}{\stype}{\type}$
  for each valuation $\rho$.
\end{lemma}

\begin{lemma}
  If $\EnvRelMatch{\Delta}{\stype}{\type}$
  then $\codeTrType{\stype} = \codetupIII{P}{\Delta'}{\type'}$
  for some $P$, $\Delta'$, $\type'$.
  Moreover, there exists valuation $\rho$ and substitution $\theta$
  such that $\RhoRelSubType{\Delta}{\cdot}{\bind{\theta}{\type'}}{\type}$
  and $\RhoRelSubType{\Delta}{\cdot}{\type}{\bind{\theta}{\type'}}$.
\end{lemma}

\begin{lstlisting}[language=meta,mathescape=true]
subtype($\type_0$; $\type_0'$) =
  match ($\type_0$; $\type_0'$) with
  | ($\alpha$; $\alpha'$) =>
    if $\alpha = \alpha'$ then ($\varnothing$; $\ftop$) else fail
  | ($\tforallT{\alpha} \type$; $\tforallT{\alpha} \type'$) =>
    let ($\Omega$, $\form$) = subtype($\type$, $\type'$) in
    ($\Omega$; $\form$)
  | ($\tforallE{\alpha} \type$; $\tforallE{\alpha} \type'$) =>
    let ($\Omega$, $\form$) = subtype($\type$, $\type'$) in
    ($\Subst{\alpha}{\effpure}{\Omega}$; $
      \form \land \Omega \toFormula{\alpha}$)
  | ($\type_1 \tarrow{\effect} \type_2$; $\type_1' \tarrow{\effect'} \type_2'$) =>
    let ($\Omega_1$, $\form_1$) = subtype($\type_1'$, $\type_1$) in
    let ($\Omega_2$, $\form_2$) = subtype($\type_2$, $\type_2'$) in
    ($\Omega_1 \cup \Omega_2 \cup \{\effect \subtp \effect'\}$; $
      \form_1 \land \form_2$)
  | _ => fail
\end{lstlisting}

To ensure soundness and completeness of this procedure
we prove the following two lemmas.
\begin{lemma}
  If $\codesubtype{\type_1}{\type_2} = \codetupII{\Omega}{\form}$
  for types $\type_1$, $\type_2$ defined over $\Delta$,
  then $\RhoRelSubType{\Delta}{\Omega}{\type_1}{\type_2}$
  for every valuation $\rho$ satisfying $\form$.
\end{lemma}

\begin{lemma}
  If $\RhoRelSubType{\Delta}{\Omega}{\type_1}{\type_2}$,
  then $\codesubtype{\type_1}{\type_2} = \codetupII{\Omega'}{\form}$
  for some $\Omega'$ and $\form$ such that
  $\rho\models\form$ and $\RhoRel{\Delta}{\Omega}{\Omega'}$.
\end{lemma}


\begin{lstlisting}[language=meta,mathescape=true]
norm($\alpha \subtp \effect$; $\form$) = $
  \{ \alpha \effguard \form \subtp \effect \}$
norm($\effpure \subtp \effect$; $\form$) = $\varnothing$
norm($\effect_1 \effjoin \effect_2 \subtp \effect$; $
  \form$) = norm($\effect_1 \subtp \effect$; $\form$) $
  \cup$ norm($\effect_2 \subtp \effect$; $\form$)
norm($\effect_1 \effguard \formAlt \subtp \effect$; $
  \form$) = norm($\effect_1 \subtp \effect$; $\formAlt \land \form$)
\end{lstlisting}

\begin{lstlisting}[language=meta,mathescape=true]
normalize($\Omega$) = $\bigcup_{(\subeffect{\effect_1}{\effect_2})\in\constrs}$ norm($\subeffect{\effect_1}{\effect_2}$; $\ftop$)
\end{lstlisting}

\begin{lstlisting}[language=meta,mathescape=true]
separate($\Delta_g$; $\Omega$) =
  let $\Omega_n$ = normalize($\Omega$) in
  let $\Omega_g$ = $\{ \gamma \effguard \form \subtp \effect \mid
    (\gamma \effguard \form \subtp \effect) \in \Omega_n \land
    \gamma \in \Delta_g \}$ in
  let $\Omega_p$ = $\{
    \beta \effguard \form \subtp \bind{\subst{
      \overline{\gamma \in \Delta_g \mapsto \effpure}}} \effect \mid
    (\beta \effguard \form \subtp \effect) \in \Omega_n \land
    \beta \notin \Delta_g \}$ in
  ($\Omega_g$; $\Omega_p$)
\end{lstlisting}

\begin{lstlisting}[language=meta,mathescape=true]
inst($\sforall{\Delta}{\Omega} \tau$) = ($\Delta$; $\Omega$; $\tau$)
\end{lstlisting}

\begin{lstlisting}[language=meta,mathescape=true]
infer($\Gamma$; $\varX$) =
  let ($\sforall{\Delta_1}{\Omega_1} \type$) = $\Gamma(\varX)$ in
  ($\varnothing$; $\Delta_1$; $\type$; $\effpure$; $\Omega_1$; $\ftop$)
\end{lstlisting}

\begin{lstlisting}[language=meta,mathescape=true]
infer($\Gamma$; $\lam{\varX}{\stype} e$) =
  let ($P_1$; $\Delta_1$; $\type_1$) = tr_type($\stype$) in
  let ($P_2$; $\Delta_2$; $\type_2$; $\effect$; $\Omega$; $\form
    $) = infer($\Gamma, \varX : \type_1$; $e$) in
  ($P_1 \uplus P_2$; $\Delta_1 \uplus \Delta_2$; $
    \type_1 \tarrow{\effect} \type_2$; $\effpure$; $
    \Omega$; $\form$)
\end{lstlisting}

\begin{lstlisting}[language=meta,mathescape=true]
infer($\Gamma$; $\lamT{\alpha} e$) =
  let ($P_1$; $\Delta_1$; $\type$; $\effect$; $\Omega_1$; $\form_1
    $) = infer($\Gamma$; $e$) in
  ($P_1$; $\Delta_1$; $\tforallT{\alpha} \type$; $\effpure$; $
    \Omega_1 \cup \{\effect \subtp \effpure\}$; $\form_1$)
\end{lstlisting}

\begin{lstlisting}[language=meta,mathescape=true]
infer($\Gamma$; $\lamE{\alpha} e$) =
  let ($P_1$; $\Delta_1$; $\type$; $\effect$; $\Omega_1$; $\form_1
    $) = infer($\Gamma$; $e$) in
  fresh $P_s$ = $\{ \fvarP_\beta \mid \beta \in \Delta_1 \}$ in
  fresh $\Delta_1'$ = $\{ \gamma_\beta \mid \beta \in \Delta_1 \}$ in
  let $\theta$ = $\subst{
    \overline{\beta \in \Delta_1 \mapsto \gamma_\beta
      \effjoin \alpha \effguard \fvarP_\beta}}$ in
  let $\Omega_1'$ = $\bind{\theta}
    (\Omega_1 \cup \{\effect \subtp \effpure\})$ in
  ($P_1 \uplus P_s$; $\Delta_1'$; $\tforallE{\alpha} \bind{\theta}\type
      $; $\effpure$; $
    \Subst{\alpha}{\effpure}{\Omega_1'}$; $
    \form_1 \land \Omega_1' \toFormula{\alpha}$)
\end{lstlisting}

\begin{lstlisting}[language=meta,mathescape=true]
infer($\Gamma$; $e_1\;e_2$) =
  let ($P_1$; $\Delta_1$; $\type_1$; $\effect_1$; $\Omega_1$; $\form_1
    $) = infer($\Gamma$; $e_1$) in
  match $\type_1$ with
  | $\type_a \tarrow{\effect} \type_v$ =>
    let ($P_2$; $\Delta_2$; $\type_2$; $\effect_2$; $\Omega_2$; $\form_2
      $) = infer($\Gamma$; $e_2$) in
    let ($\Omega_s$, $\form_s$) = subtype($\type_2$; $\type_a$) in
    ($P_1 \uplus P_2$; $\Delta_1 \uplus \Delta_2$; $
      \type_v$; $\effect_1 \effjoin \effect_2 \effjoin \effect$; $
      \Omega_1 \cup \Omega_2 \cup \Omega_s$; $
      \form_1 \land \form_2 \land \form_s$)
  | _ => fail
\end{lstlisting}

\begin{lstlisting}[language=meta,mathescape=true]
infer($\Gamma$; $e\;[\stype]$) =
  let ($P_1$; $\Delta_1$; $\type_1$; $\effect_1$; $\Omega_1$; $\form_1
    $) = infer($\Gamma$; $e$) in
  match $\type_1$ with
  | $\tforallT{\alpha} \type$ =>
    let ($P_2$; $\Delta_2$; $\type_2$) = tr_type($\stype$) in
    ($P_1 \uplus P_2$; $\Delta_1 \uplus \Delta_2$; $
      \Subst{\alpha}{\type_2}{\type}$; $\effect_1$; $
      \Omega_1$; $\form_1$)
  | _ => fail
\end{lstlisting}

\begin{lstlisting}[language=meta,mathescape=true]
infer($\Gamma$; $e\;[\seffect]$) =
  let ($P_1$; $\Delta_1$; $\type_1$; $\effect_1$; $\Omega_1$; $\form_1
    $) = infer($\Gamma$; $e$) in
  match $\type_1$ with
  | $\tforallE{\alpha} \type$ =>
    let ($\Delta_2$; $\effect_2$) = tr_effect($\seffect$) in
    ($P_1$; $\Delta_1 \uplus \Delta_2$; $
      \Subst{\alpha}{\effect_2}{\type}$; $\effect_1$; $
      \Omega_1$; $\form_1$)
  | _ => fail
\end{lstlisting}

\begin{lstlisting}[language=meta,mathescape=true]
infer($\Gamma$; {let} $x$ = $e_1$ {in} $e_2$) =
  let ($P_1$; $\Delta_1$; $\type_1$; $\effect_1$; $\Omega_1$; $\form_1
    $) = infer($\Gamma$; $e_1$) in
  fresh $\Delta_1'$ = $\{ \beta_\alpha \mid \alpha \in \Delta_1 \}$ in
  fresh $\Delta_g$ = $\{ \gamma_\alpha \mid \alpha \in \Delta_1 \}$ in
  let $\theta$ = $\subst{\overline{\alpha \in \Delta_1
    \mapsto \beta_\alpha \effjoin \gamma_\alpha}}$ in
  let ($\Omega_g, \Omega_p$) = separate($\Delta_g$; $
    \bind{\theta}(\Omega_1 \cup \{\effect_1 \subtp \effpure\})$) in
  let ($P_2$; $\Delta_2$; $\type_2$; $\effect_2$; $\Omega_2$; $\form_2
    $) = infer($
      \Gamma, x : \sforall{\Delta_g}{\Omega_g} \bind{\theta}\type_1$; $e_2$) in
  ($P_1 \uplus P_2$; $\Delta_1' \uplus \Delta_2$; $\type_2$; $\effect_2$; $
    \Omega_p \cup \Omega_2$; $\form_1 \land \form_2$)
\end{lstlisting}

\end{document}